 \documentclass[preprint,review,10pt,sort&compress]{elsarticle}
\usepackage[margin=1.18 in]{geometry} %1.1 10pts
\usepackage{multirow}
\usepackage{adjustbox}
\usepackage{graphicx}
\usepackage{gensymb}
\usepackage{xcolor}
\usepackage{dirtytalk}
\usepackage[title]{appendix}
\newcommand{\bc}{\begin{center}}
\newcommand{\ec}{\end{center}}
\newcommand{\bfr}{\begin{flushright}}
\newcommand{\efr}{\end{flushright}}
\newcommand{\ii}{\item}
\newcommand{\be}{\begin{enumerate}}
\newcommand{\ee}{\end{enumerate}}
\newcommand{\bi}{\begin{itemize}}
\newcommand{\ei}{\end{itemize}}
\newcommand{\bd}{\begin{description}}
\newcommand{\ed}{\end{description}}
\newcommand{\beq}{\begin{equation}}
\newcommand{\eeq}{\end{equation}}
\newcommand{\bea}{\begin{eqnarray}}
\newcommand{\eea}{\end{eqnarray}}

\newcommand{\bfi}{\begin{figure}}
\newcommand{\efi}{\end{figure}}
\newcommand{\bay}{\begin{array}{l}}
\newcommand{\eay}{\end{array}}

\newcommand{\etal}{\textit{et al.}}

\usepackage{amsmath}
\usepackage{breqn}
\usepackage{textcomp}
\usepackage[flushleft]{threeparttable}
\usepackage{booktabs}
\usepackage{caption}
\usepackage{makecell}
\usepackage{tabularx}
\definecolor{UW}{RGB}{64, 38, 96}
\usepackage{amssymb}
\journal{Composite Part A: Manufacturing}

\begin{document}

%%%%%%%%%%%%%%%%%%%%%%%%%%%%%%%%%%%%%%%%%%%%%%%%%%%%%%%%%%%%%%%%%%%%%
%%%%%%%%%%%%%%%%%% START FRONT PAGE WITH UW LOGO %%%%%%%%%%%%%%%%%%%%
%%%%%%%%%%%%%%%%%%%%%%%%%%%%%%%%%%%%%%%%%%%%%%%%%%%%%%%%%%%%%%%%%%%%%

\begin{titlepage}
\clearpage\thispagestyle{empty}
%\noindent {\footnotesize {{\em

%\hfill To be submitted to Cement and Concrete Composites} }} \\
\noindent
\hrulefill

\begin{figure}[h!]
\centering
\includegraphics[width=1.5 in]{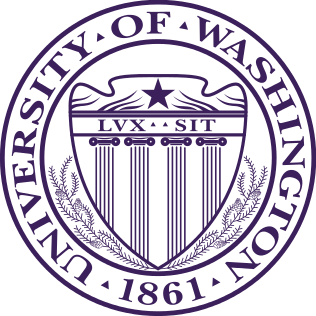}
\end{figure}
%{\color{NU} some text}

\begin{center}
{\color{UW}{{\bf A\&A Program in Structures} \\ [0.1in]
William E. Boeing Department of Aeronautics and Astronautics \\ [0.1in]
University of Washington \\ [0.1in]
Seattle, Washington 98195, USA}}
\end{center} %\vskip 5mm

\hrulefill \\ \vskip 2mm
\vskip 0.5in

\begin{center}
{\large {\bf Effects of Average Number of Platelets Through the Thickness and Platelet Width on the Mechanical Properties of Discontinuous Fiber Composites}}\\[0.5in]
{\large {\sc Seunghyun Ko, Troy Nakagawa, Zhisong Chen, William B. Avery, Ebonni J. Adams, Matthew R. Soja, Michael H. Larson, Chul Y. Park, Jinkyu Yang, Marco Salviato}}\\[0.75in]
{\sf \bf INTERNAL REPORT No. E-0223}\\[0.75in]
\end{center}

\noindent {\footnotesize {{\em Submitted to Composite Part A: Manufacturing \hfill February 2023} }}
\end{titlepage}

\newpage
%%%%%%%%%%%%%%%%%% END FRONT PAGE WITH UW LOGO %%%%%%%%%%%%%%%%%%%%
%%%%%%%%%%%%%%%%%%%%%%%%%%%%%%%%%%%%%%%%%%%%%%%%%%%%%%%%%%%%%%%%%%%

\begin{frontmatter}



\title{Effects of Average Number of Platelets Through the Thickness and Platelet Width on the Mechanical Properties of Discontinuous Fiber Composites}

% \author[label1]{Seunghyun Ko}
\author[label1]{Seunghyun Ko}
\author[label1]{Troy Nakagawa}
\author[label1]{Zhisong Chen}
\author[label1]{William B. Avery}
\author[label2]{Ebonni J. Adams}
\author[label3]{Matthew R. Soja}
\author[label3]{Michael H. Larson}
\author[label4]{Chul Y. Park}
\author[label1]{Jinkyu Yang}
\author[label1]{Marco Salviato \corref{cor1}}
\address[label1]{William E. Boeing Department of Aeronautics and Astronautics, University of Washington, Seattle, WA 98195, USA}
\address[label2]{Boeing Defense, Space, and Security, Huntsville, AL 35808, USA}
\address[label3]{Boeing Commercial Airplanes, Boeing Company, Renton, WA 98057, USA}
\address[label4]{Supernal LLC, Washington DC, 20036, USA}

\cortext[cor1]{Corresponding Author, \ead{salviato@aa.washington.edu}}

%%%%%%%%%%%%%%%%%%%%%%%%%%%%%%%%%%%%%%%%%%%%%%%%%%%%%%%%%%%%%%%
%%%%%%%%%%%%%%%%%%%%%%%%%%%%%%%%%%%%%%%%%%%%%%%%%% ABSTRACT
%%%%%%%%%%%%%%%%%%%%%%%%%%%%%%%%%%%%%%%%%%%%%%%%%%%%%%%%%%%%%%%
\begin{abstract}
\linespread{1}\selectfont
In this study, we experimentally and numerically investigate the evolution of the tensile material properties of Discontinuous Fiber Composites (DFCs) with an increasing average number of platelets through the thickness for two different platelet widths. %We test $30$ coupons of $0.15"$ and $0.25"$ thicknesses, and $15$ coupons of $0.065"$ thickness to have statistically significant data. MS I do not think this is necessary in the abstract

The results show that both the number of platelets and the platelet width have significant effects on the tensile modulus and strength. We find that not only the average mechanical properties but also their coefficients of variation change according to the different DFC mesostructures. 

To understand the relationship between material morphology at the mesoscale and corresponding material properties, we developed a random platelet mesostructure generation algorithm combined with explicit finite element models. Leveraging the computational tools, we find that moduli and strength increase with increasing average number of platelets through the thickness. The increasing trend continues until reaching an asymptotic limit at about $45$ layers through the thickness for the narrow platelets and $27$ layers for the square platelets.

In the study, we address the importance of having accurate simulations of the mesostructure to match not only the average modulus and strength but also their associated coefficients of variation. We show that it is possible to accurately predict the tensile material properties of DFCs, including their B-basis design values. This is a quintessential condition for the adoption of DFCs in structural applications.

\end{abstract}

\begin{keyword}
Discontinuous reinforcement \sep Statistical properties/methods \sep Computational Modelling \sep Mechanical properties

\end{keyword}

\end{frontmatter}

%%%%%%%%%%%%%%%%%%%%%%%%%%%%%%%%%%%%%%%%%%%%%%%%%%%%%%%%%%%%%%%
%%%%%%%%%%%%%%%%%%%%%%%%%%%%%%%%%%%%%%%%%%%%%%%%%% Introduction
%%%%%%%%%%%%%%%%%%%%%%%%%%%%%%%%%%%%%%%%%%%%%%%%%%%%%%%%%%%%%%%
\section{Introduction}
Several industrial sectors including aerospace, automotive, and wind energy have seen an increasing demand of lightweight and strong structures made of carbon fiber reinforced polymers (CFRPs) \cite{Barbero2018, Tsai2014, Elmarakbi2013, Chortis2013}. The rising demand is pushing the manufacturing of CFRPs to the limit of the current technologies in terms of productivity and costs. As the successful example of the Boeing 787 clearly proves \cite{Boeing2021}, traditional continuous fiber composites are suitable for large and relatively simple contour structures such as wings and fuselages \cite{Baker2004}. However, because of fiber folding and wrinkling, continuous fiber composites are not an ideal choice to manufacture components featuring very complex shapes such as hinges, brackets, and other joint structures \cite{Rashidi2018,Wang2015} such is the case especially in the context of the ever increasing push in the aerospace industry for enhanced production rate and cost effectiveness. Using a discontinuous fiber form is a promising alternative approach to manufacture such complex structures whose manufacturing has been traditionally dominated by metals.  

Discontinuous Fiber Composites (DFCs) or chopped fiber composites are composed of randomly deposited platelets or chips made from uni-directional fiber prepregs chopped into a defined shape and size. In the literature, DFCs and their variants have also been called Randomly Oriented Strands (ROS) \cite{Sel2016}, Stochastic Prepreg Platelet Molded Composites (PPMC) \cite{Denos2018a}, Composite Oriented Strand Board \cite{Jin2017}, Tow-Based Discontinuous Composites (TBDCs) \cite{Li2017}, Ultra-Thin Chopped carbon fiber Tape reinforced Thermoplastic (UT-CTT) \cite{Wan2016a}, Sheet Molding Compound (SMC) composites \cite{Chen2018}, and Stochastic Tow-Based Discontinuous Composite (STBDC) \cite{Ryatt2022}. 

The platelets can be molded into complex contours thanks to their discontinuity which makes DFCs a suitable material for compression molding and, in case of thermoplastic DFCs, injection molding. In terms of their mechanical properties, DFCs provide superior performance compared to short fiber composites. In fact, while short fiber composites feature fiber volume fractions below $30\%$ \cite{Fatemi2015}, DFCs maintain $50\%$ or higher fiber volume fractions \cite{Boursier2001, Aubry2001}. This allows DFCs to maintain a tensile stiffness near $80\sim90\%$ of quasi-isotropic laminates made of identical prepregs and a tensile strength comparable to aluminum \cite{Visweswaraiah2018}. As a result, DFCs represent excellent candidate materials to address the demand for high-rate and cost-efficient production of CFRP structures with complex contours and high performance requirements.

Platelet-based DFCs provide significant advantages in terms of manufacturing but the price to pay is a more complex mechanical behavior compared to traditional composites. In fact, DFCs feature a more complex inhomogeneous mesostructure which makes their behavior strongly dependent on the average orientation of the platelets, the platelet shape and size, the average number of platelets through the thickness, and the type of matrix. Furthermore, recent studies highlighted the emergence of significant size effects on the structural strength \cite{Ko2019a, Ko2019b, troy22}. %Therefore, characterizing the meso-structures become the major challenge. 

Several researchers studied the relationship between mesostructure characteristics and mechanical properties. Feraboli \etal~\cite{Fera2009a,Fera2009b} investigated the effects of platelet size and coupon thickness on elastic properties and strengths. They found that longer platelets and thicker structures led to better mechanical performance. Later, Selezneva \etal~\cite{Sel2016} expanded the study finding that also the presence of local warpage, out-of-plane platelet orientations, and resin-rich areas impacted the mechanical performance of DFCs. Wan and Takahashi~\cite{Wan2016a} investigated the link between manufacturing parameters and mechanical performance of DFCs. They found that increasing hot-press pressure decreases the void contents therefore increasing the tensile strength while the modulus did not show particular correlation with the pressure. Li \etal~\cite{Li2017} found that thinner platelets can lead to higher tensile strength compared to standard platelets for a given shape and size. Ko \etal~\cite{Ko2019a, Ko2019b} found that the mode I fracture energy is a function of the platelet size and the average number of layers through the thickness with longer platelets and thicker coupons providing higher fracture energies. Furthermore, they noticed that DFCs feature a much larger fracture process zone \cite{Baz1998,Sal21a,Baz2004,Baz2017,Yao2020a,Yao2018a,Yao2018b,Yao2018c, Baz1996, Sal2019, Sal2016a, Li21, Sal2016b, Brockmann2023} compared to continuous fiber composites and other quasibrittle materials. The fracture process zone in front of the notch was found to be $3\sim7$ times larger than in continuous fiber composites denoting a very significant damage tolerance of DFCs structures compared to traditional composites. 

Among many different mesostructure characteristics, the platelet orientation has drawn significant attention. To explicitly measure the platelet orientation, researchers started to utilize X-ray micro Computer Tomography ($\mu$CT). Explicitly capturing the individual fibers using the $\mu$CT scan was challenging due to the limited scan resolution. Instead, each voxels contained homogenized density of fiber and matrix. Denos \etal~\cite{Denos2014, Denos2018a,Denos2017a}, Kravchenko~\cite{Kravchenko2017}, Favaloro \etal~\cite{Favaloro2018} and later Wan and Takahashi~\cite{Wan2016b, Wan2018, Wan2021}, measured the platelet orientation tensors using the density gradient between the neighboring voxels obtained from the $\mu$CT images. They assumed that the platelet is in the direction of the least density difference between the voxels while the normal direction was set to be the direction of largest density difference. Their results agreed with microscope images in which they visually confirmed the fiber orientation. Furthermore, they integrated the platelet orientations into finite element models and analyzed complex 3D structures~\cite{Denos2018a, Kravchenko2017}. Each finite elements contained platelet orientations measured by $\mu$CT scans capturing the flow of the platelets such as swirling around the corners. The numerical models accurately predicted failure modes of the DFC structures. Kravchenko~\cite{Kravchenko2017} found that when there is no significant platelet directional flow in the manufacturing process, the average orientation tensor remained close to a perfectly random distribution. On the other hand, Martulli \etal~\cite{Martulli2019} analyzed DFC coupons with directional flow which induced significant platelet orientation change compared to the no-flow coupons.

Based on the numerous experimental data, computational models of DFCs have been developed through the years to understand the relationship between the mesostructure characteristics and mechanical properties. Feraboli \etal~\cite{Fera2010} and Harban \etal~\cite{Tuttle2017} introduced discretized, partition-based stochastic finite element models for flat coupons. They divided the 2D space into square partitions and stored unique platelet orientations generated by a certain probability distribution. This partition-based approach was further improved by Selezneva \etal~\cite{Sel2017} who explicitly generated platelets over the partitioned space, stacking them to create the final mesostructure allowing to effectively capture the random strain field in DFCs. Ko \etal~\cite{Ko2018a,Ko2018b,Ko2020,Ko2021}, Kravchenko \etal~\cite{kravchenko2019a}, Sommer \etal~\cite{Sommer2020a}, Shah \etal~\cite{Shah2020}, Visweswaraiah \etal~\cite{Visweswaraiah2017}, Chen \etal~\cite{Chen2018}, and Ryatt \etal~\cite{Ryatt2022} contributed to improving the explicit generations of the platelets by introducing additional features such as (a) algorithms for the prevention of excessive platelet concentrations, (b) thickness adjustment strategies to account for matrix flow and platelet deformation, and (c) approaches to account for out-of-plane orientations. Other researchers also worked on analytical methods. Li and Pimenta \cite{Li2019} and Henry and Pimenta~\cite{Henry2017}, for instance, developed a semi-analytical method equipped with shear-lag analysis to predict the strength of DFCs and the effect of platelet waviness. The model captured the dependence of the tensile strength on the platelet thickness. Wan \etal~\cite{Wan2021} developed an analytical approach combined with the partition-based platelet generation to find the modulus of flat DFC coupons. The model matched the experimental data for different platelet sizes.

Several studies showed a good agreement between computational model predictions and average experimental results. However, very few investigations considered the prediction of the Coefficient of Variation (CoV) of such properties. Without matching property variations, engineers and designers must rely on a high margin of safety, leading to overdesign and material waste. Only when the models successfully match both average and variation, DFCs can be safely utilized.

In this paper, we study the effect of the coupon thickness and the platelet width on the tensile elastic modulus and strength focusing on both average and CoV of the mechanical properties. To do so, we first test a large number of coupons to create statistically significant experimental data. In fact, DFCs possess high scatter in experimental data compared to continuous fiber composites \cite{Fera2009a,Fera2009b,Sel2016,Wan2016a,Li2017,Ko2019a, Ko2019b,Kravchenko2017,kravchenko2019a}. Due to the high variation, the number of test coupons must be larger than the minimum number of recommended test coupons (typically $5$) suggested by ASTM D3039 test standard for continuous fiber composites. Unfortunately, most of the experimental investigations available in the open literature report no more than $5-8$ repetitions per test except for Kravchenko \etal~\cite{kravchenko2019a} who, to address the importance of statistical variations, tested $10\sim24$ specimens per case. However, a comprehensive investigation on the effects of layer numbers and platelet size on the mechanical properties, where a statistically meaningful number of repetitions is used in all the tests, is still elusive. We present it in the following sections of the article. Another important activity performed in this work is computationally generating the mesostructures focusing on local platelet orientation variations. We implement the spatial distributions of the platelets including their Coefficients of Variation (CoV). By precisely controlling the CoV, we create more realistic mesostructures allowing us to study the statistical distribution of mechanical properties.

%%%%%%%%%%%%%%%%%%%%%%%%%%%%%%%%%%%%%%%%%%%%%%%%%%%%%%%%%%%%%%%
%%%%%%%%%%%%%%%%%%%%%%%%%%%% (2) EXPERIMENT
%%%%%%%%%%%%%%%%%%%%%%%%%%%%%%%%%%%%%%%%%%%%%%%%%%%%%%%%%%%%%%%
\section{Experiments}
\subsection{Materials and methods}
We investigated three coupon thicknesses and two platelet morphologies: square and narrow. The square platelet is $12.7 ~\text{mm} \times 12.7 ~\text{mm}$ ($0.5" \times 0.5"$) while the narrow platelet has an identical length but significantly reduced width (the exact dimension is intellectual property and cannot be disclosed). We tested coupon thicknesses of $1.65~\text{mm}~(0.065"),~3.81~\text{mm}~(0.15"),~\text{and}~ 6.35~\text{mm}~(0.25")$ for the narrow platelets and $1.65~\text{mm}~(0.065"),~\text{and}~3.81~\text{mm}~(0.15")$ for the square platelets. These thicknesses roughly correspond to $11$, $27$, and $45$ layers through the thickness on average. %We will represent the coupon thickness using the average number of layers through the thickness in the following sections. 

Sekisui Aerospace~\cite{Sekisui} manufactured $305~\text{mm} \times 305~\text{mm}$ flat DFC plates made of AS4D/PEKK thermoplastic tapes from Solvay. We discarded $12.7$ mm from the edges, and trimmed $254~\text{mm} \times 25.4~\text{mm}$ tensile coupons using a wet tile saw with a diamond-coated blade to trim the plates. Each coupon had two $50.8$ mm garolite tabs at the ends, leaving $152.4$ mm as a gauge length. After manufacturing the samples, We sprayed a layer of flat white paint with black paint speckles on a single side to measure the surface displacements using the Digital Image Correlation (DIC) technique. Then, we tested the coupons using a ShoreWestern servo-hydraulic axial-torsion load frame with a closed-loop control and $250$ kN capacity. The loading strain rate was set at 0.6$\%$ strain per minute for a total duration of the test of $2\sim3$ minutes per coupon. A digital camera captured the surface with a sampling rate of 1 Hz. Taken images were analyzed using GOM software \cite{Gom} to measure the surface displacements.

We carefully selected the number of tested coupons to fully capture the statistics of DFCs. Cochran's sample size formula \cite{Cochran1977} was used to determine the sample population to meet desired accuracy from a given probability: 
\begin{equation}\label{eqCochran}
n = \left[\frac{Z*\sigma}{PA*\mu-\mu}\right]^2
\end{equation}
where $n$ is the sample size, Z is the confidence level set equal to 95$\%$ ($=1.96$), $\sigma$ is the data standard deviation, $\mu$ is the data average value, and $PA$ (Percent Accuracy) is the desired accuracy set to $95\%$ in this work. For the data average and standard deviation, we used uni-axial tension test results from Kravchenko et al. \cite{kravchenko2019a} where they tested $24$ coupons. The average strength ($\mu$) was $228.8$ MPa and the standard deviation ($\sigma$) was $27.7$ MPa. Using Eq.~\ref{eqCochran}, the suggested sample size ($n$) was $22$ which, as can be noticed is much larger than the traditional sample size ($5 \sim 8$) suggested for continuous composites. Hence, we tested more than $30$ coupons to fully characterize the statistics of DFCs. For the case of $11$ layers through the thickness ($1.65$ mm), we tested only $15$ coupons due to manufacturing difficulties involving thin panels. However, the sample size was still higher than the traditional sample size. It is worth noting that typical DFC structures feature thicknesses of at least 18 layers ($2.54$ mm) to avoid warpage and to reduce the amount of voids and resin rich areas in the structures. Hence, the choice of investigating a thickness starting from 11 layers ($1.65$ mm) was made mostly to ensure our investigation also covered extreme design cases.

\subsection{Experimental Results} \label{Sec-ExperimentalResults}
\begin{figure}
 \centering
  \includegraphics[trim=0cm 0cm 0cm 0cm, clip=true,width = 1\textwidth, scale=1]{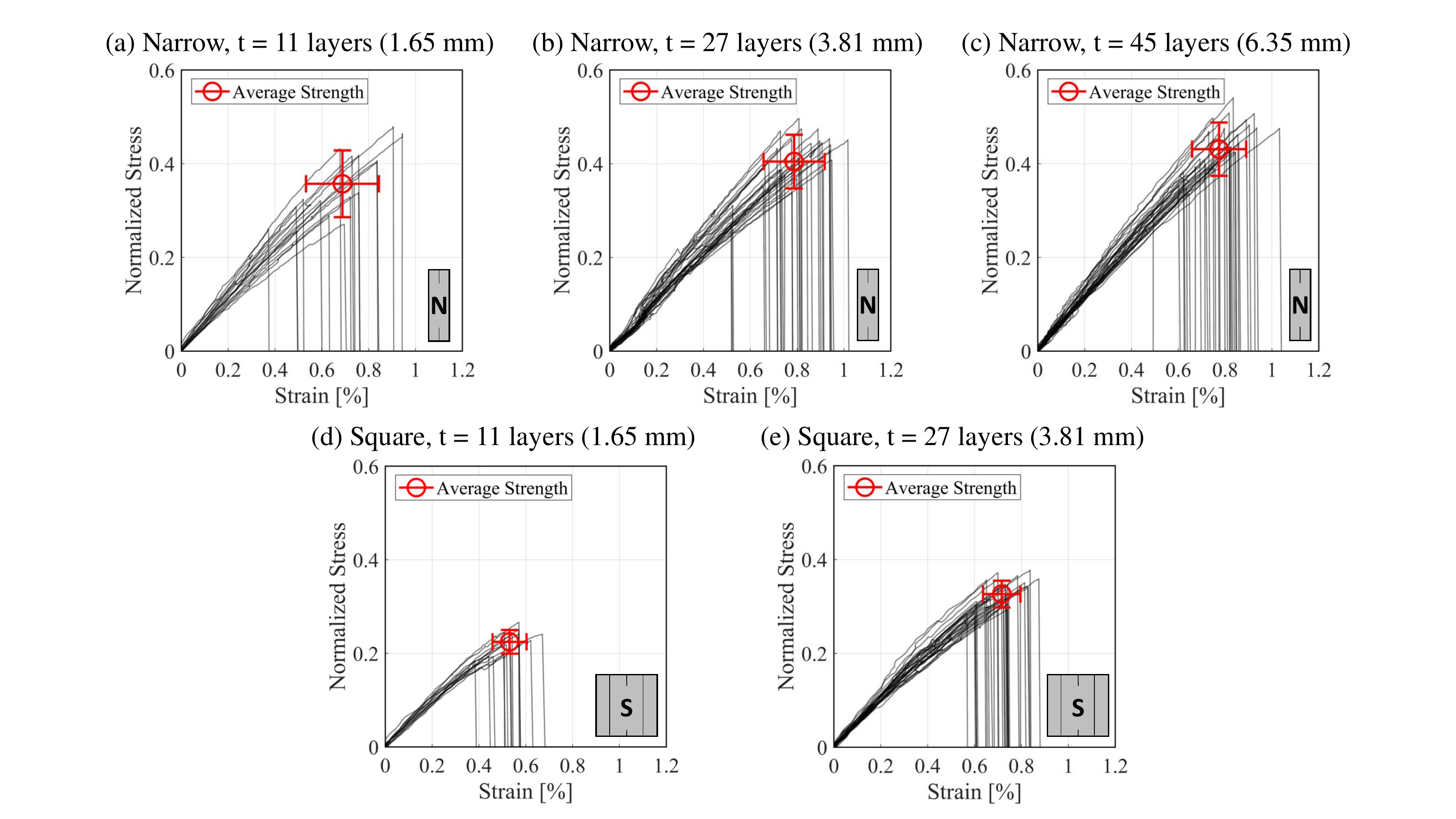} \caption{\label{f-TestSigEpsCurves} Experimental normalized stress and strain curves for coupons of various thicknesses made with narrow (a-c) and square platelets (d,e). The tested DFC coupons show strong variability in the tensile modulus and strength. The mean strength is plotted as a circle with one standard deviation using the error bars. Compared to the case of narrow platelets, the samples made with square platelets exhibit significantly lower variation ($12.9\%$ vs. $6.7\%$ at the thickness of 27 layers) in their tensile strength.} 
\end{figure}

For all the platelet widths and average number of layers through the thickness, we can observe significant scatter in the tensile elastic modulus and strength. In Fig.~\ref{f-TestSigEpsCurves}, normalized stress and strain curves of the narrow and square platelets with different thicknesses are plotted where the stress is normalized against the theoretical strength of a quasi-isotropic laminate made of the same prepreg utilized to fabricate the DFC panels. For specimens made of narrow platelets with 45 layers through the thickness ($6.35$ mm), the percent difference between the minimum and the maximum strength is nearly $52\%$. We checked for outliers in the test data using the maximum normed residual method suggested by the Composite Material Handbook-17 (CMH-17) \cite{CMH2002}. All the test data safely passed the outlier test. A summary of the test results is listed in Table~\ref{T1}. 

\begin{figure}
 \centering
  \includegraphics[trim=0cm 2cm 0cm 0cm, clip=true,width = 1\textwidth, scale=1]{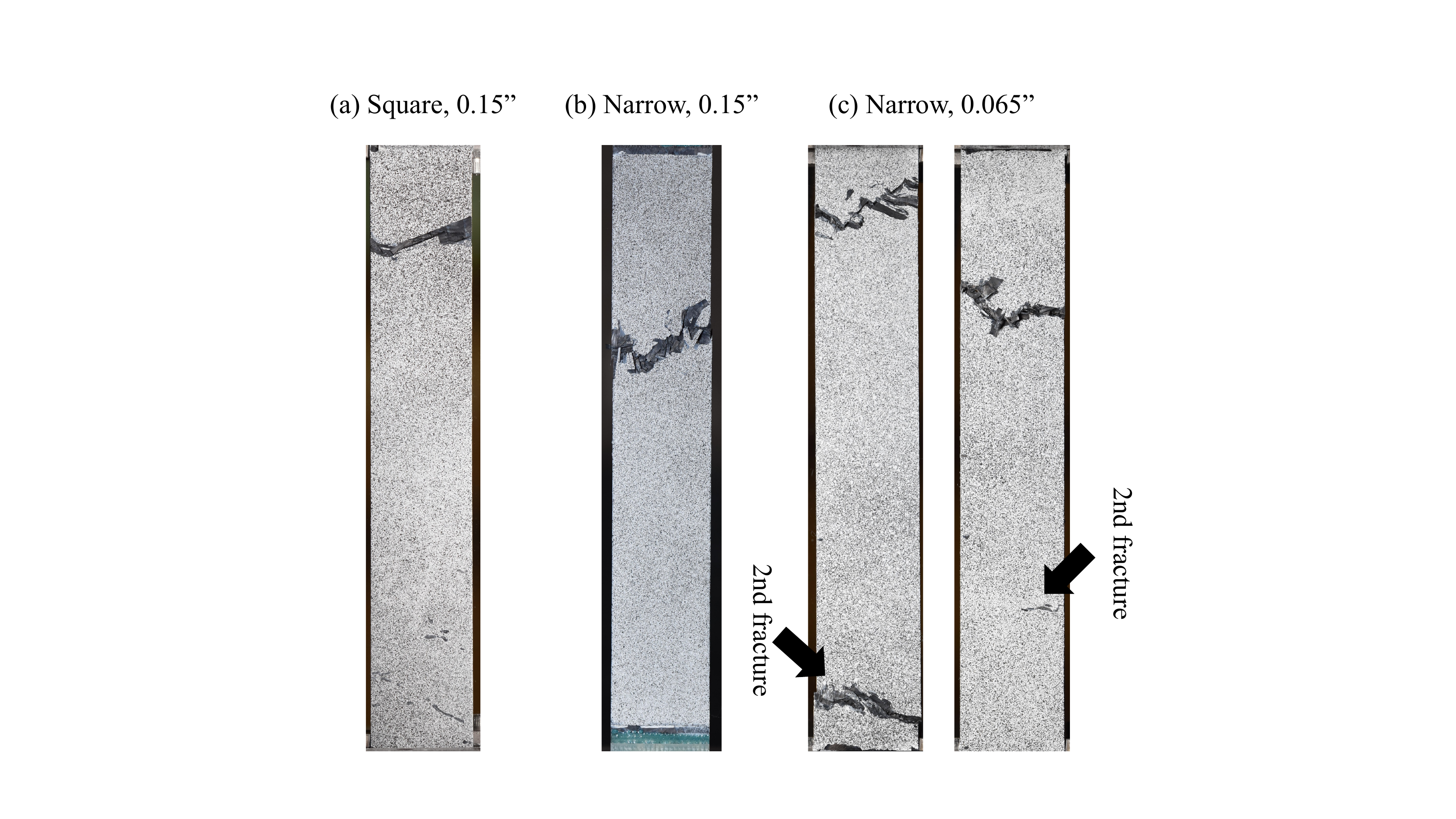} \caption{\label{f-FractureSurface} Representative fracture surfaces of the tested DFC coupons. The specimens with narrow platelets exhibit more tortuous fracture surfaces compared to the coupons with square platelets. %Also, we observe two coupons with double fractures from the narrow platelets with $0.065$" thickness. However, the strength shows no significant difference compared to the rest of the group.
  } 
\end{figure}

Highly random fracture patterns are observed for all the coupons as can be noted in Fig.~\ref{f-FractureSurface} which shows four representative fracture surfaces for different platelet widths and thicknesses. Interestingly, for the $11$ layers ($1.65$ mm) with narrow platelets, two out of $22$ coupons failed at two distinct locations (see Fig.~\ref{f-FractureSurface}c). These double fractures happened only for the $11$ layers ($1.65$ mm) thickness with the narrow platelets. %The double fractures must be strongly related with the randomness associated with the mesostructures. 
It is worth noting that for the $11$ layers ($1.65$ mm) thick plates, we observed significant plate warpage. Selezneva et al. \cite{Sel2016} showed that the warpage increases drastically as the thickness decreases. This is caused by the paucity of platelets through the thickness which leads to highly non-symmetric, unbalanced layups. Therefore, these warped plates have strong axial-twist coupling. When the axial load was applied, strong twisting led to double fractures near the ends of the coupons. However, we did not observe any significant strength and modulus reduction due to the double fractures.

\begin{figure}
 \centering
  \includegraphics[trim=0cm 3cm 0cm 0cm, clip=true,width = 1\textwidth, scale=1]{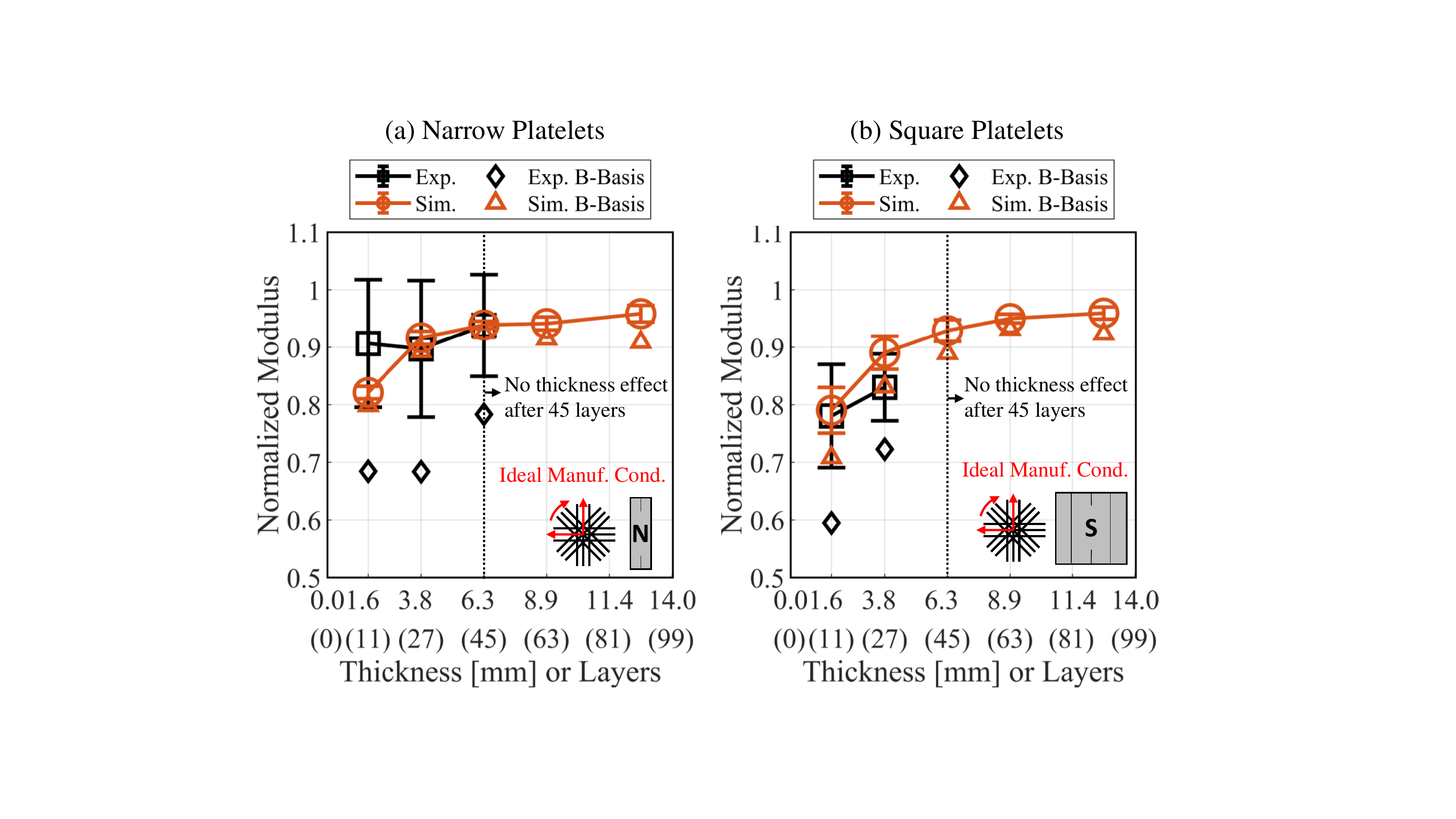} \caption{\label{f-modulus} Summary of the modulus thickness effect for the (a) narrow and (b) square platelets observed by experiments and simulations. %The thickness effect in the tensile elastic modulus is insignificant for both platelet sizes. 
  The simulation results predict that the modulus of both platelet sizes reaches $97\%$ of the quasi-isotropic laminate modulus at the thickness of $12.7$ mm corresponding to $91$ layers through the thickness.} 
\end{figure}

\begin{figure}
 \centering
\includegraphics[trim=0cm 3cm 0cm 0cm, clip=true,width = 1\textwidth, scale=1]{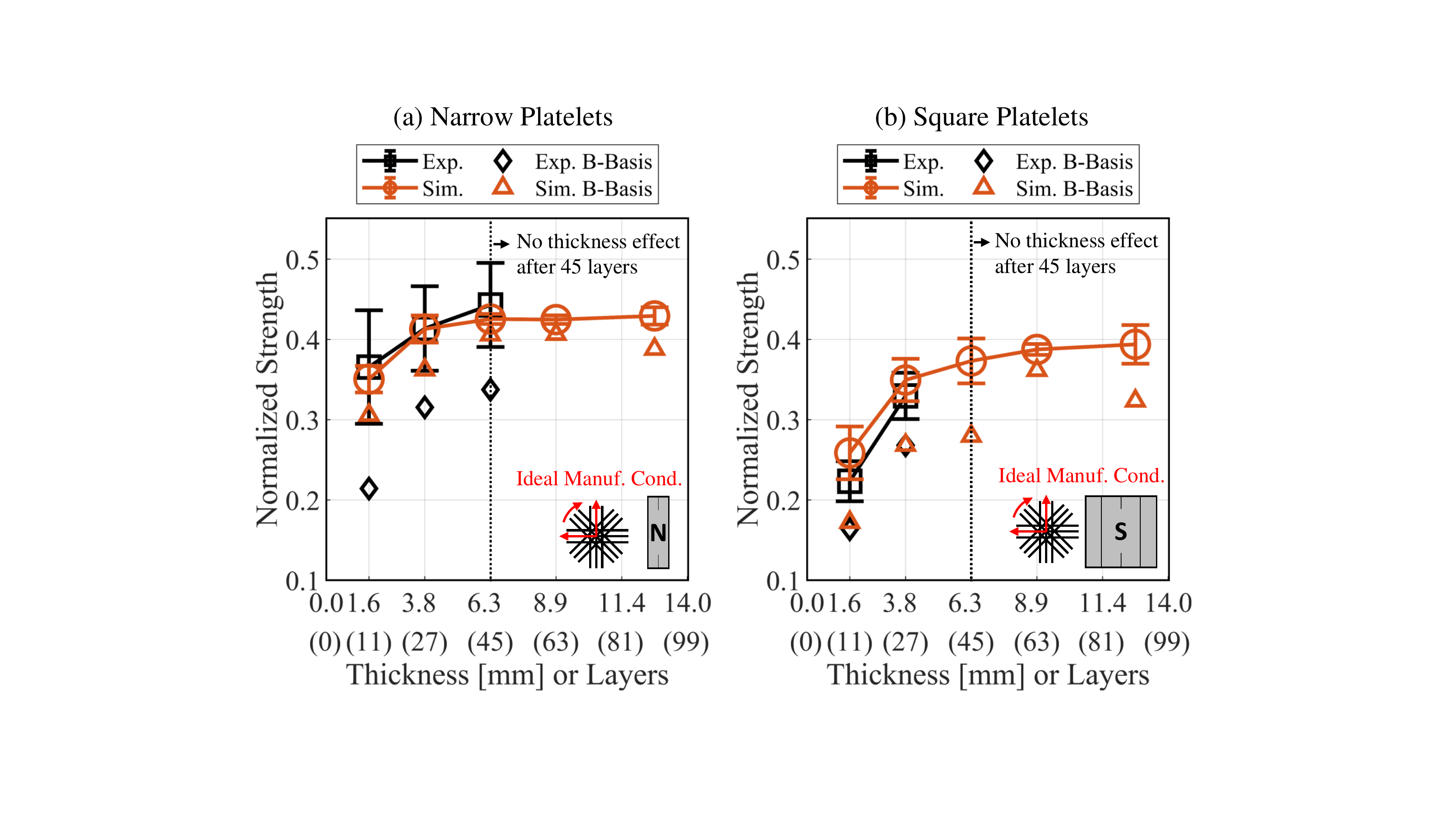} \caption{\label{f-strength} Summary of the thickness effect on the strength for the (a) narrow and (b) square platelets observed by experiments and simulations. %A significant thickness effect is observed in both platelet sizes. 
Using the simulations, we find that the strength reached the asymptotic level after a coupon thickness of $6.35$ mm corresponding to 45 layers through the thickness. At the asymptotic limit, the strength difference between the two platelets is only $7$ percent.} 
\end{figure}

The analyzed experimental results are plotted in Fig.~\ref{f-modulus} and \ref{f-strength} where tensile elastic modulus and strength are normalized against the theoretical quasi-isotropic laminate properties calculated using the Classical laminate theory and Tsai-Hill failure criterion. When the thickness increases from $11$ layers ($1.65$ mm) to $45$ layers ($6.35$ mm), the strength of the DFCs made with narrow platelets increases by $19.6\%$ and the modulus increases by $4.3\%$. For the case of square platelets, the strength and modulus are increased by $32.4\%$ and $6.0\%$ when the thickness increases from $11$ layers ($1.65$ mm) to $27$ layers ($3.81$ mm) showing that the thickness effect is more pronounced for the square platelets. Comparing the mechanical properties of the two platelet sizes, the narrow platelet outperforms the square both in terms of modulus and strength for all thicknesses. The percent differences between the two platelets are $15.1\%$ in the modulus and $48.5\%$ in the strength for the $11$ layers ($1.65$ mm) thickness. The differences decrease as the thickness increases. In section \ref{sec-SimEandSig}, we predict the modulus and the strength for a thickness of $91$ layers ($12.7$ mm) using a finite element model showing that the differences become smaller and smaller for sufficiently large number of platelets through the thickness. However, the narrow platelets still outperforms the square ones by $7.6\%$ in strength. 

The narrow platelet coupons have higher average modulus and strength throughout the thicknesses but a trade-off exists. In fact, the specimens made with narrow platelets also exhibit a higher Coefficient of Variation (CoV). The CoV decreases with increasing thickness which is a trend also found by Kravchenko \etal~\cite{kravchenko2019a} who observed that the decreasing variation in the mechanical properties is related to the decreasing local variability. As the structure thickness increases, the local orientation variation decreases because there is a larger number of platelets through the thickness. However, while this observation is certainly true, there is also another aspect that must be considered. When the platelet width decreases, despite the fact that there are more platelets in the coupon, the modulus and strength variations increase. Here, we must distinguish between the through thickness variation and the spatial variation. The through thickness variation decreases when the total number of platelets increases. On the other hand, the spatial variation, i.e. the variation of the orientation tensor as a function of the spatial location within the structure, generally increases when the total number of platelets increases. Therefore, we must be cautious when we relate the variations of the material properties to the total number of the platelets. We further investigate the spatial variation with different platelet widths in section \ref{sec-PlateletWidthEffects}. 

Besides the average and CoV of the experimental results, we also calculated B-Basis design values following CMH-17 \cite{CMH2002}. The B-Basis design value is the lower tenth percentile of the distribution with a $95\%$ confidence level. In other words, $10$ out of $100$ specimens will fail below the B-basis value with $95\%$ confidence level. To calculate the B-basis value, we first identified the probability distributions of the test data following the Anderson-Darling statistic test suggested by CMH-17 \cite{CMH2002}. It was found that the modulus data was fitted well by the normal distribution. The strength data could be fitted by both normal and the Weibull distributions. However, we chose to use the Weibull distribution as CMH-17 \cite{CMH2002} recommended for strength. At the coupon thickness of $11$ layers ($1.65$ mm), the statistic test was near fail to fit the distribution. Therefore, using the B-basis design value is not recommended below $11$ platelets through the thickness. 

The B-basis allowables are plotted as diamond markers in Fig.~\ref{f-modulus} and Fig.~\ref{f-strength}. As can be noted, the B-Basis values are strongly affected by the CoVs. Due to the strong variation in the $11$ layer thick coupons, the narrow platelet B-Basis is lowered by $41\%$ compared to the average strength. Also, a high variation in the narrow platelet results in $5.4\%$ decrease in the B-basis modulus compared to the square platelet at the thickness of $27$ layers ($3.81$ mm). However, the average modulus for the narrow platelet coupons of thickness $27$ layers ($3.81$ mm) is actually $6.7\%$ higher than the square platelet. Indeed, the material property variation plays a significant role. A list of the B-basis design values is given in Table~\ref{T2}.  

\subsection{Measuring the platelet orientation using X-ray $\textit{\micro}$CT}
%Creating the meso-structures with realistic manufacturing conditions required extra steps. First, an X-ray micro-CT scan (North Star Imaging X5000) was used to find the full 3D orientation tensors of the platelets. Each scanned area was roughly 3” by 3” and required two separate scans per coupon. The scanned images were stitched during the post-processing. The average voxel resolution of the scanned parts was roughly 25 μm. The density gradient method with a Gaussian smoothing filter was used to find the fiber orientations. It determined the principal fiber orientation along the least change of density between voxels as suggested by Denos et al. and Wan et al. [25, 26]. We averaged the orientation tensors within 1” by 1” sections, creating 6 average orientation tensors (A11) per coupon. The section size was chosen following [5]. However, depending on the platelet size and thickness, the ideal section size could change. In Figure 6, an example of CT scan image with the average orientation tensors are shown. For this coupon, section #6 corresponds to the location of the final fracture where A11 is least favorable. A33 indicates the out-of-plane orientation. Because the plate was manufactured without strong platelet flow, the A33 is significantly lower than A11 or A22. We scanned three of the 0.15” thick and five of the 0.25” thickness coupons with the narrow platelets. A summary of the average orientation tensors is listed in Table IV.

We measured the platelet orientation distribution using X-ray micro-computed tomography. Toward this goal, we scanned $5$ square platelet coupons with an average of $27$ layers through the thickness ($3.81$ mm) Using a NSI X5000 scanner~\cite{NSI} with a scan resolution of $25$ $~\micro$m while the reconstruction of the scanned images was performed via VGStudio Max \cite{VGStudio2015}. 

Several methods to calculate the orientation tensors of platelets from $\micro$CT scans have been proposed in the literature \cite{Wan2016b,Favaloro2018,Denos2017a,Denos2017b,Denos2018a,Martulli2019}. We chose to adapt the approach by Denos \etal~\cite{Denos2017a,Denos2017b,Denos2018a} which calculates the density gradient between the neighboring voxels and correlates it with the platelet orientation tensors. We used 3D Gaussian distribution derivative filters to find the local density gradients while a Gaussian smoothing filter was used to reduce the noise~\cite{Shapiro2011}.

%%%%%%%%%%%%%%%%%%%%%%%%%%%%%%%%%%%%%%%%%%%%%%%%%%%%%%%%%%%%%%%
%%%%%%%%%%%%%%%%%%%%%%%%%%%% (3) Computational Modeling
%%%%%%%%%%%%%%%%%%%%%%%%%%%%%%%%%%%%%%%%%%%%%%%%%%%%%%%%%%%%%%%

\section{Computational Modeling Framework}
We developed a computational modeling framework to understand the effects of the mesostructure characteristics on the material properties of DFCs. The model follows chiefly three steps: 
\begin{enumerate}[leftmargin=2cm]
    \item[\textbf{Step 1}: ] Perform random platelet mesostructure generation,
    \item[\textbf{Step 2}: ] Assign random platelet spatial variation, and 
    \item[\textbf{Step 3}: ] Perform stochastic finite element simulations of the mechanical behavior.
\end{enumerate}
The first and second steps create random mesostructures while the last step transforms the mesostructures into finite element models with proper constitutive laws and damage formulations. 
%%% (3-1) 
\subsection{Random platelet meso-structure generation}
The mesostructure made of randomly deposited platelets governs the mechanical behaviors of DFCs. In fact, it has been shown extensively that the platelet length, aspect ratio, thickness, and orientations influence the mechanical performances of DFCs \cite{Fera2009a,Fera2009b,Sel2016,Wan2016a,Li2017,Ko2019a, Ko2019b,Kravchenko2017,kravchenko2019a}. Therefore, we prioritize generating mesostructures by closely mimicking the manufacturing process. We follow the manufacturing process called mat-stacking method introduced in \cite{Jin2017, Ko2018b, Ko2019a}.

The main purpose of the mesostructure generation algorithm is to store unique platelet information explicitly. We adopt the discretized, partition-based method \cite{Fera2010,Tuttle2017,Steiner1994} where the individual platelet information is stored into partitions whose size, determined to have at least three subdivisions across the smallest length of the platelet, is $0.51~\text{mm} \times 0.51~\text{mm}$ ($0.02"\times0.02"$). Our code can explicitly generate platelets with user-defined geometries and orientations. In the present work, since we verified that the platelets did not experience high flow conditions, we use a uniform random orientation distribution for the platelet generation although the algorithm could easily accommodate any probability distribution (e.g., Gaussian, or trapezoidal). %We deposit the platelets into random locations inside the $305~\text{mm} \times 305~\text{mm}$ space. %The partitions interfere with the platelets store the orientations. 

During the platelet generation, multiple platelets can be deposited in a partition but we keep track of the order of the deposition. Therefore, through the thickness, the platelets are deposited in a ``tetris-like" fashion while the algorithm controls the number of platelets in each partition. In fact, when platelets are generated without restrictions, a significant uneven distribution of the platelets is often observed, as similarly reported by Selezneva \etal~\cite{Sel2017}. To prevent such extreme unevenness, we developed a saturation point method \cite{Ko2019a, Ko2019b} where the number of assignable platelets is forced to saturate at every $n$ number of platelets through the thickness. We choose $n = 3$ because the amount of a single $305~\text{mm} \times 305~\text{mm}$ mat that we create during the manufacturing process is roughly three platelets through the thickness. The saturation point method controls the variation in the number of platelets through the thickness and prevents unrealistic local paucity or concentration of the platelets. 

The generation algorithm stops when the average number of platelets through the thickness reaches the target value. %After the generation, the platelets through the thickness still vary locally. 
Then, we adjust the thickness of the platelets and introduce resin-rich layers to mimic the actual hot-press process where the coupon thickness is adjusted by the resin and fiber flow. Basically, when the local thickness passes the target thickness, the individual platelet thickness is evenly reduced. On the other hand, when the local thickness fails to reach the target, we introduce resin-rich layers. It is worth mentioning that the introduced resin-rich layers occupy less than $5\%$ of the coupon volume which is in very good agreement with experimental measurements showing volume fractions in the order of $5\sim10\%$ \cite{Sel2016, Sommer2020b}. 

\begin{figure}
 \centering
\includegraphics[trim=0cm 0cm 0cm 0cm, clip=true,width = 1\textwidth, scale=1]{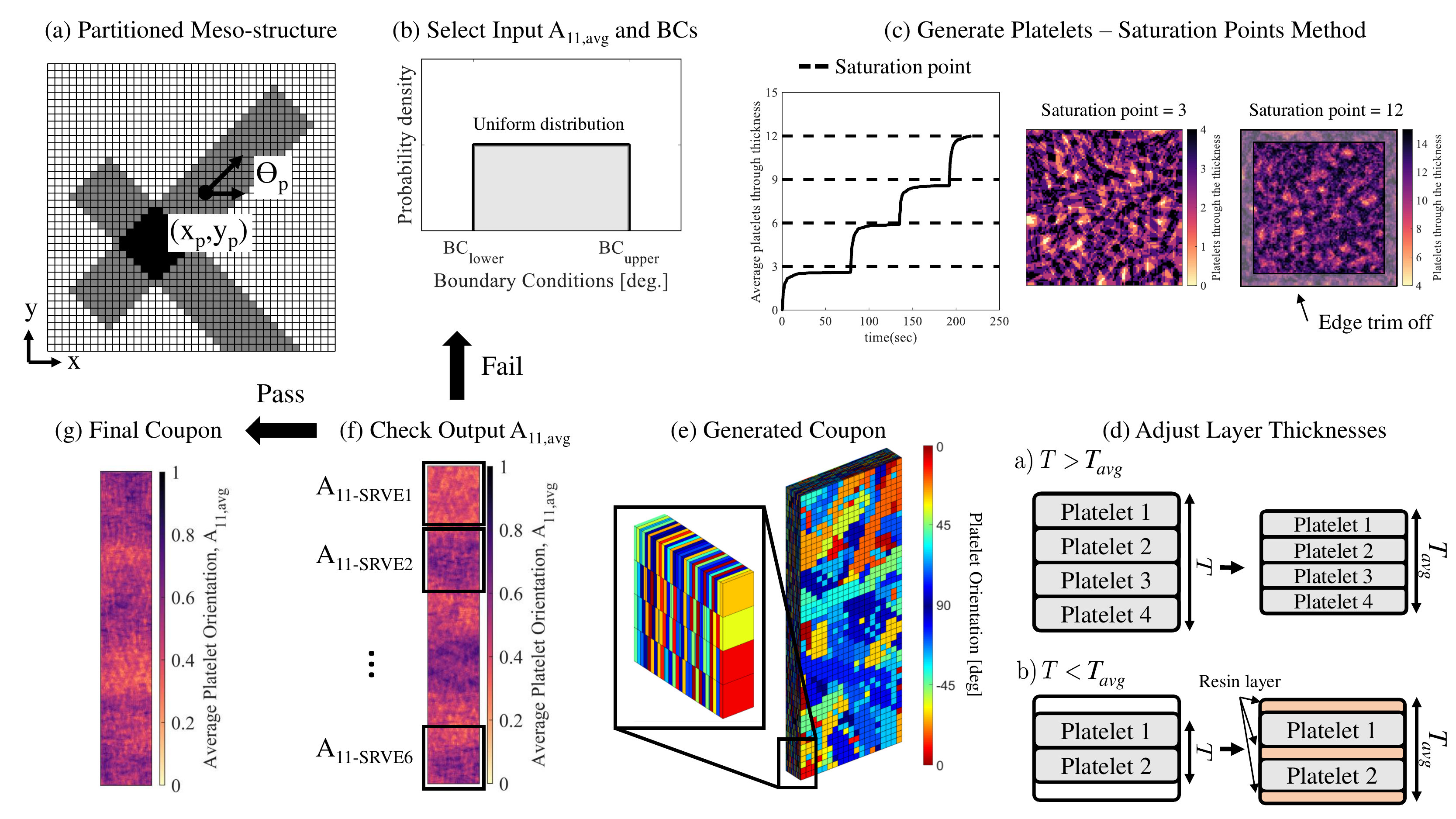} \caption{\label{f-MesoGenerator} Graphical flowchart describing the random platelet mesostructure generation. Using the generation algorithm, we control both average and variation of the explicitly generated platelet orientations. Detailed information can be found in ~\cite{Ko2019a,Ko2019b,Ko2018a,Ko2018b,Ko2020,Ko2021}} 
\end{figure}

After the thickness adjustment, we trim at about 12.7 mm from the edges to discard edge effects. In Fig.~\ref{f-MesoGenerator}, a road map of the random platelet meso-structure generation is plotted. The generated meso-structures are carefully designed to mimic the manufacturing process and carry all the deposited platelet information. It is worth noting that we have extensively validated the algorithm in past studies. Figure ~\ref{f-MesoGenerator} will be revisited in the later sections, and more information about this algorithm can be found in \cite{Ko2019a,Ko2019b,Ko2018a,Ko2018b,Ko2020,Ko2021}.

%%% (3-2) 
\subsection{Random platelet in-plane spatial variation}
In addition to the mesostructure generation algorithm, we implemented a random platelet spatial distribution to precisely control the statistical variation of the platelet orientation. %It controls the platelet orientation locally varying in space. 
In DFCs, there are two important platelet orientation variations: through the thickness and in-plane spatial. The saturation point method implemented in the random platelet meso-structure generation controls the variation through the thickness. Here, we will explain how we control the variation in space using the spatial variation algorithm.

% In Fig.~\ref{f-modulus} and ~\ref{f-strength}, we observe the effects of the coupon thickness (which is related to the average number of layers through the thickness) and platelet width. It can be noted that the coupon thickness effect is more related with the variation through the thickness while the platelet width effect is more closely related to the variation in space {\color{red} What makes you say this? It is not clear. The plots only show that there is an effect of the thickness and of the width. They do not discriminate between spatial or thickness variations MS}.They are not independent from each other but one has a stronger impact than the other.

\begin{figure}
 \centering
\includegraphics[trim=4cm 6cm 4cm 4cm, clip=true,width = 1\textwidth, scale=1]{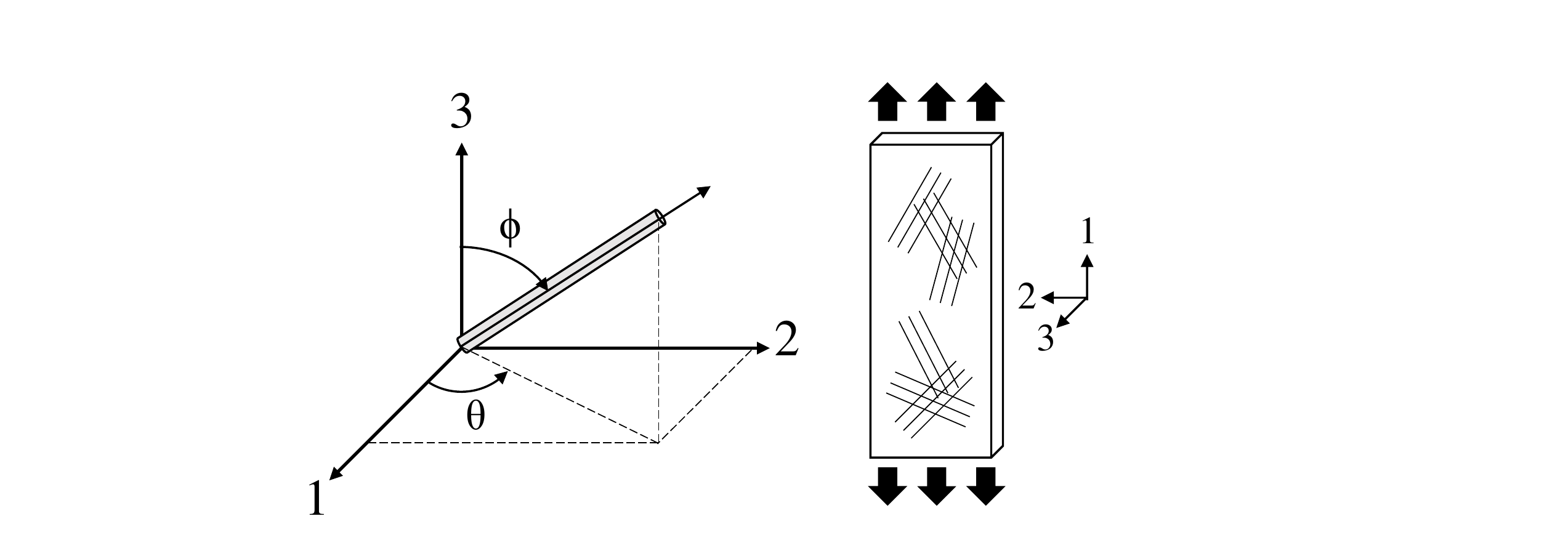} \caption{\label{f-FiberOrientation} Coordinate system and definitions of the in-plane angle, $\theta$, and the out-of-plane angle, $\Phi$~\cite{Tucker1987}. The direction $1$ corresponds to the loading direction of the investigated coupons. The direction $3$ is aligned with the thickness direction.} 
\end{figure}

First, we study what is the actual fiber orientation variation over the space. We define the 3D platelet orientation tensor, $A_{ij}$, following \cite{Tucker1987}:
\begin{equation}\label{eqOrientTensor}
A_{ij} = 
\begin{bmatrix}
A_{11} &A_{12}  &A_{13} \\ 
... & A_{22} & A_{23}\\ 
sym & ...  & A_{33}
\end{bmatrix}
= 
\begin{bmatrix}
 p_1p_1& p_1p_2 &p_1p_3 \\ 
... &  p_2p_2&p_2p_3 \\ 
sym & ... & p_3p_3
\end{bmatrix}
\end{equation}
where $p_1 = sin(\phi)cos(\theta)$, $p_2 = sin(\phi)sin(\theta)$, and $p_3 = cos(\phi)$ and the in-plane angle, $\theta$, and the out-of-plane angle, $\phi$, are defined as shown in Fig.~\ref{f-FiberOrientation}. In the figure, direction 1 is the longitudinal direction (loading direction), direction 2 is the transverse direction, and the in-plane angle $\theta$ is positive from direction 1 to 2. The out-of-plane angle $\phi$ is positive from direction 3 to plane 1-2. The diagonal terms, $A_{ii}$ (summation not implied), can take values from 0 to 1 and their sum is equal to 1. 

Once the orientation tensors are defined, we calculate the average orientation through the thickness using the following equation:
\begin{equation}\label{eqOrientTensorAverage}
A_{ii,avg}=\frac{1}{N\cdot t_{tot}}\sum_{j=1}^{N}A_{ii}\cdot t_j
\end{equation}
where $n$ is the number of the platelets through the thickness excluding resin-rich layers. 

The average values of the orientation tensor, $A_{ii,avg}$, give an idea of the average orientation of the platelets. When the platelets are oriented following a uniform random distribution in all three directions, $A_{11,avg} = A_{22,avg} = A_{33,avg} = 1/3$. On the other hand, when the platelets remain flat (2D case) and follow a uniform distribution for all the direction in plane 1-2, $A_{11,avg} = A_{22,avg} = 1/2$ and $A_{33,avg} = 0$. 

We have utilized X-ray $\micro$CT scans to measure the fiber orientations. Following Eq.~\ref{eqOrientTensorAverage}, we averaged the fiber orientations of $5$ square platelet coupons and found: $A_{11,avg} = 0.511$, $A_{22,avg} = 0.486$, and $A_{33,avg} = 0.003$. As can be noted, the $A_{33,avg}$ was negligible compare to other directions. A similar result was found by Kravchenko \textit{et al}. \cite{Kravchenko2017} who CT scanned $10$ DFC coupons and found that $A_{33,avg} = 0.007$. Therefore, we assumed $A_{33,avg}$ to be negligible. Since the summation of $A_{11,avg}$ and $A_{22,avg}$ is equal to $1$, we can refer to the 2D average orientation tensor using a single parameter $A_{11,avg}$. If all the platelets are oriented towards the loading direction, $A_{11,avg} = 1$. If they are oriented towards the transverse direction, $A_{11,avg} = 0$.

After calculating all the average components of the orientation tensor, we divide a coupon into several regions and enforce unique platelet orientations in each division. Each division represents a Statistical Representative Volume Element (SRVE) and has a size of $25.4~\text{mm} \times 25.4~\text{mm}$ \cite{Denos2017a, Kravchenko2017}. The SRVE serves as the smallest unit of the statistical sample size which still retains all the main statistical characteristics of the population. The SRVE considers the statistical fluctuations of the effective properties over the finite domain \cite{Kanit2003}. %Each SRVE possesses different mesostructures but the grouped SRVEs represent the heterogeneous DFC structures. 
Therefore, understanding the effective properties of SRVEs such as $A_{11,avg}$ is a critical step to understand the statistical variations of the DFC coupons. %However, we acknowledge that the statistics of $A_{11,avg}$ depends on the size of SRVEs. We will further investigate how to select the SRVE size in future studies.

We found the distributions of $A_{11,avg}$ between the SRVEs of the square platelet coupons using the $\micro$CT scans. The mean $A_{11,avg}$ is $0.49$ with a CoV of $9.1\%$. The maximum $A_{11,avg}$ is $0.556$ and the minimum $A_{11,avg}$ is $0.394$. As can be noted, the platelet orientations vary in space. Therefore, we need a tool to enforce the $A_{11,avg}$ other than $0.5$.

\begin{figure}
 \centering
\includegraphics[trim=0cm 4cm 0cm 0cm, clip=true,width = 1\textwidth, scale=1]{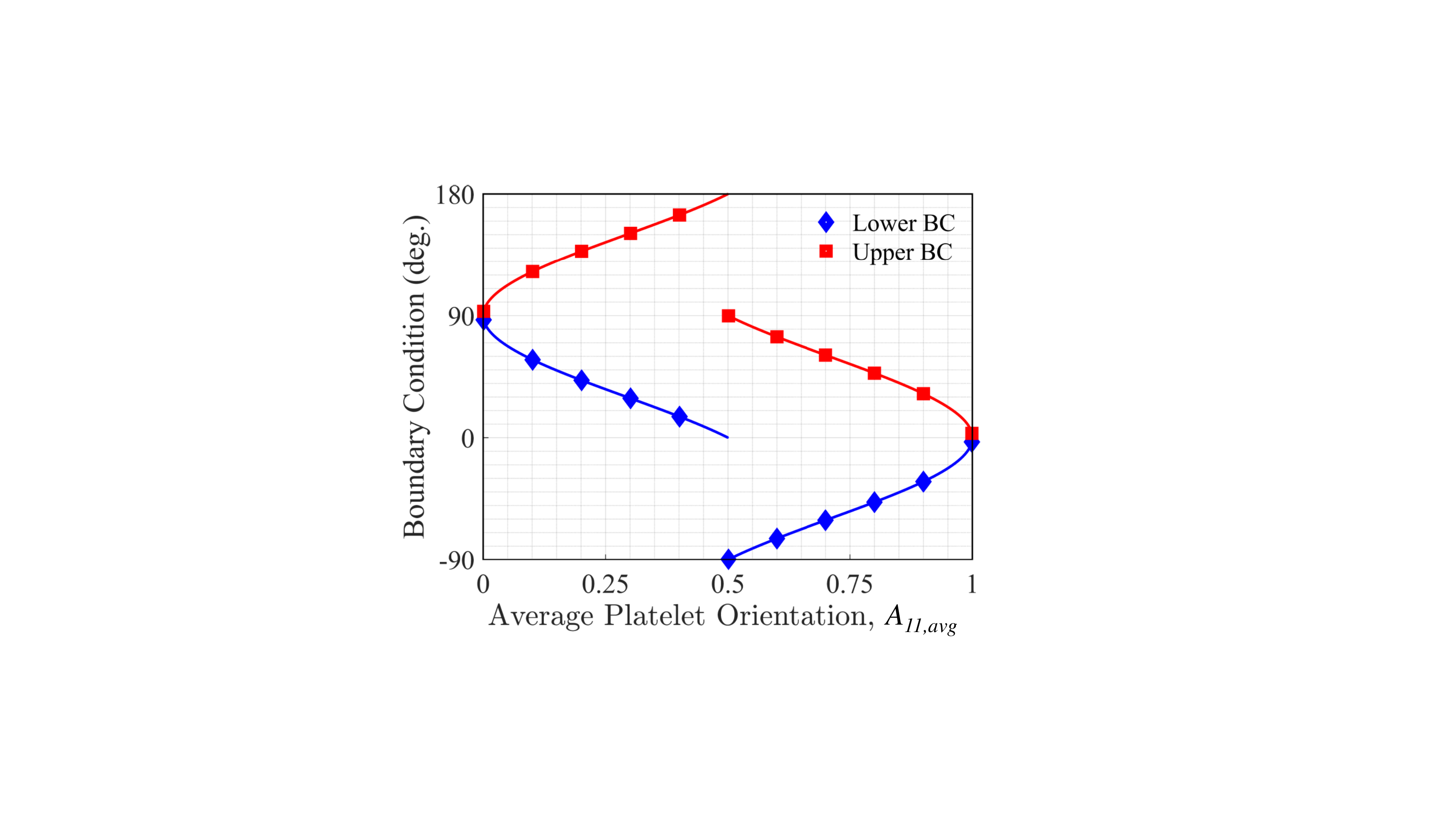} \caption{\label{f-BoundaryCondition} Lower and upper boundary conditions of the uniform distribution used to generate the platelets to match a target $A_{11,avg}$.} 
\end{figure}

In order to enforce unique $A_{11,avg}$ in each SRVE, we manipulate the boundaries of the uniform distribution which generate random platelet orientations. We can express the relationship between the $A_{11,avg}$ and the uniform distribution boundaries using the following simple integral form:  

\begin{equation}\label{eqBC}
A_{11,avg}=\frac{1}{BC_U-BC_L}\int_{BC_L}^{BC_U}cos^2(\theta)d\theta
\end{equation}
where $-\pi/2 \leq BC_{L} \leq \pi/2$ and $0 \leq BC_{U} \leq \pi$. From the foregoing equation, we find boundary conditions ($BC_{U}, BC_{L}$) that provide the target $A_{11,avg}$.  For example, if $A_{11,avg} = 0.5$, then $BC_L = -\pi/2$ and $BC_U = \pi/2$ which correspond to the ideal manufacturing condition where platelets are perfectly uniformly distributed. Figure~\ref{f-BoundaryCondition} shows the boundary conditions used to generate the $A_{11,avg}$ for general platelet distributions using Eq. (\ref{eqBC}).

The uniform probability distribution is sufficient to represent the platelet orientation distribution for the low flow coupons. The low flow coupons are manufactured without preferential flow of the platelets. The lack of preferential flow limits the $A_{11,avg}$ between $0.35$ and $0.65$ \cite{Kravchenko2017}. However, as the $A_{11,avg}$ moves toward the extremes ($0$ or $1$), the error between the actual distribution and the simulated meso-structure gets amplified. In fact, in the simulated meso-structure, the selected boundary conditions omit a portion of the platelet orientations in order to match the $A_{11,avg}$. However, in reality, the platelet orientation must fully cover $0$ to $\pi$. Therefore, if ones need to create mesostructures with high flow condition, the probability distribution of the platelet generation function needs to be selected carefully \cite{Chen2018}.

The target $A_{11,avg}$ in each region is not always easy to achieve because of the overlapping platelets between SRVEs. %It gets harder to match the target $A_{11,avg}$ as the platelet width decreases. 
A simple solution would be to generate the SRVEs separately and combine them to create a coupon. However, this method would create sharp discontinuities between the SRVEs which is unrealistic. Accordingly, we decided to generate the entire coupon at once. To accommodate the effect of the overlapping platelets, we implemented a bisection optimization algorithm. The optimization algorithm checks the difference between the current $A_{11,avg}$ and the target $A_{11,avg}$ in each SRVE. When the current $A_{11,avg}$ is lower than the target $A_{11,avg}$, we increase the input $A_{11,avg}$ by $0.01$ until the difference flips the sign. We also reduce the increment by half. We iterate the process until the difference between the target and current $A_{11,avg}$ is less than $1\%$ in all the SRVEs simultaneously. 

Equipped with the spatial variation algorithm, the random platelet mesostructure generator can control the platelet orientation variation through the thickness and over the space. A complete graphical flowchart of the mesostructure generation is plotted in Fig.~\ref{f-MesoGenerator}.

%%% (3-3) 
\subsection{Stochastic finite element framework}
\begin{figure}
 \centering
\includegraphics[trim=0cm 0cm 0cm 0cm, clip=true,width = 1\textwidth, scale=1]{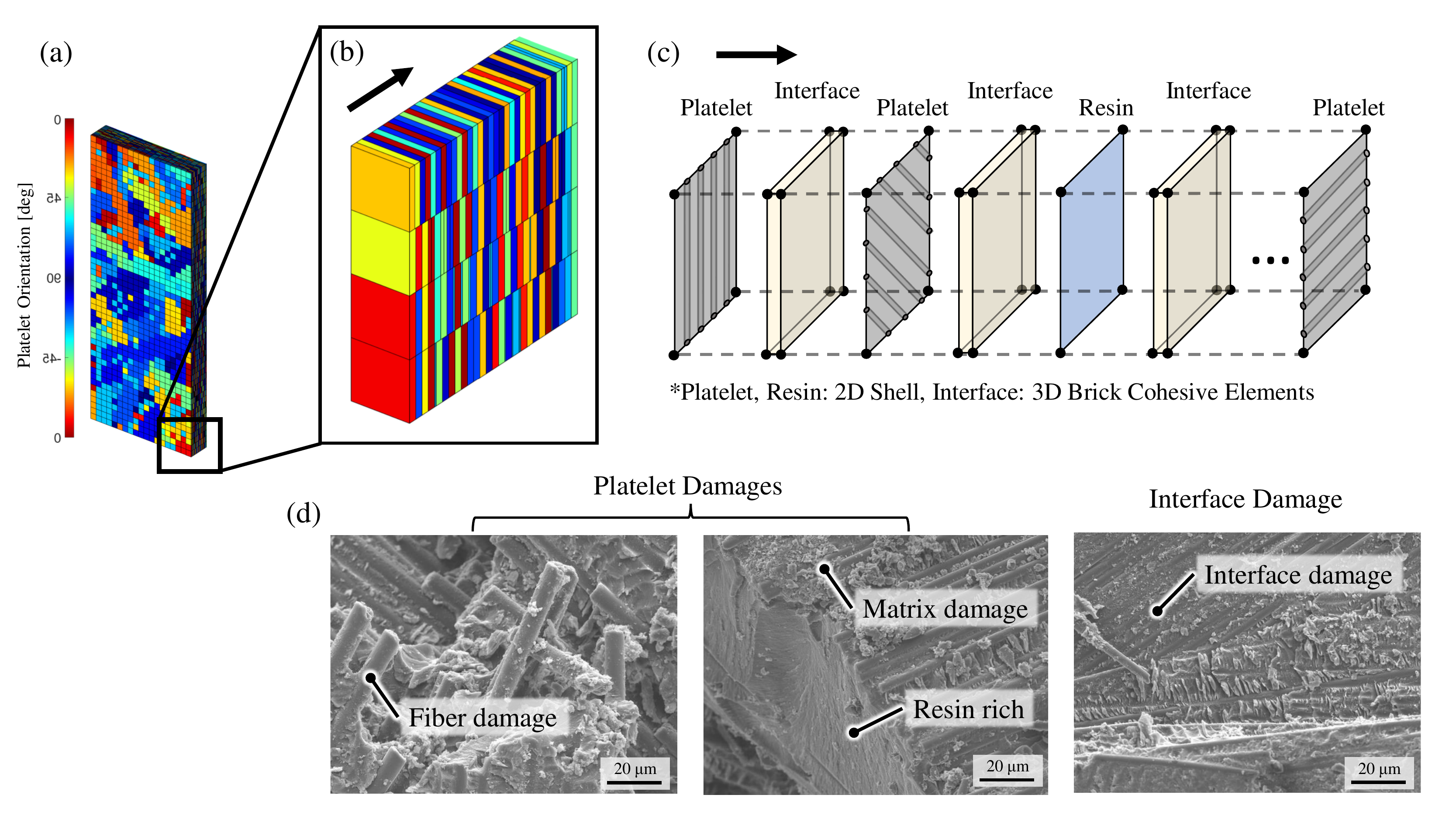} \caption{\label{f-FEMFramework} A summary of the stochastic finite element model framework. From the random platelet meso-structures (a - b), individual layers are modeled using 2D shell elements (c). Between the shell elements, the cohesive elements are inserted to represent the interface layers. Using this approach, we are able to capture three main failure mechanisms that we observed from SEM (d).} 
\end{figure}

Once the mesostructure is generated, we use the information on the platelet orientation to generate finite element models in Abaqus~\cite{Abaqus}. As shown in Fig.~\ref{f-FEMFramework}, which provides a schematic description of the FE model, we aim at capturing the intra-, and inter-laminar failure mechanisms in DFC coupons. Intra-laminar fracture involves failure mechanisms within the platelets such as fiber breakage, fiber pull-outs, and matrix damage. Often this damage starts to accumulate prior to ultimate failure. %Micro-cracks accumulate and form a macro-crack leading to the structural failure. 
Therefore, the damage process in quasibrittle materials such as CFRPs \cite{Baz1996, Sal2016a, Sal2016b, Green2007, Ko2019a, Ko2019b,  Sal21a, troy22, Jeremy22}, concrete \cite{Baz1984, Baz1990, Baz1998}, ceramics, rocks, sea ice, and nanocomposites \cite{Sal2011, Sal2013, Yao2018a, Yao2018b, Yao2018c, Sal2019} is complex and progressive.

To capture the intra-laminar damage, we use a continuum damage mechanics approach where the damage is smeared over the element and the material degradation is modeled by strain softening constitutive laws. We use Hashin failure criteria for damage initiation \cite{Hashin1980,Lapczyk2007}, and linear damage softening laws regularized according to the crack band model for the progression of the damage \cite{Baz1983,Sal21a}. The platelets are simulated using Belytschko-Tsai shell elements with reduced integration (S4R element in the Abaqus). Each partition from the mesostructures is treated as a shell element containing unique fiber orientation and thickness. It is worth mentioning here that more advanced constitutive laws for the platelets such as the \textit{Spectral Stiffness Microplane Model} \cite{Beghini2008a, Beghini2008b, Sal2016c} or the \textit{Cylindrical Microplane Model} \cite{xue2022cylindrical, sabounchi2022microplane} could be used. However, for the particular cases investigated in this work, we did not notice significant differences between the results obtained via Hashin's failure criterion and other models. 

For the inter-laminar damage, we use a cohesive zone model with a linear traction-separation constitutive behavior \cite{Camanho2002}. For damage initiation, we use the quadratic stress failure criterion while for damage progression a linear softening law with Benzeggagh-Kenane (BK) mixed mode formulation featuring an exponent of 2 is used \cite{Benzeggagh1996}. Towards this goal, a layer of 3D, 8-node cohesive elements (COH3D8 in Abaqus) with thickness of $1/100^{th}$ of the prepreg ply thickness is inserted between the shell elements. A single side of the shell elements and the cohesive elements share common nodes, creating a laminate. Between the laminates, a perfect bonding constraint is applied to create a coupon. 

The platelet material properties used in the simulations were taken from \cite{kravchenko2019b} in which almost the same material system was tested. For the cohesive element properties, the penalty stiffnesses in normal and two shear directions are chosen to be $1.0\times10^4 ~\text{N/mm}^3$ \cite{Turon2007}. The strengths are equal to $200$ MPa in all three directions as well. The normal and shear fracture energies are $1.4$ and $2.4$ N/mm. The cohesive material properties are used to calibrate the model to match the experiment results from the narrow platelets with $27$ layers ($3.81$ mm) thickness. Then, the cohesive properties are used for all the predictions of the other thicknesses and platelet sizes investigated in this work. Uni-axial displacement is applied at the end of the specimen while the other end is fixed in all directions. A mesh size of $0.51~\text{mm} \times 0.51~\text{mm}$ is used to match the partition size. All the simulations were performed in ABAQUS/Explicit leveraging an explicit integration algorithm.

% In the shell elements, we use the transverse shear stiffness ten times larger than the recommended values \cite{Abaqus} to match the linear elastic stiffness.

%%%%%%%%%%%%%%%%%%%%%%%%%%%%%%%%%%%%%%%%%%%%%%%%%%
%% (4) Computational Analysis Results
%%%%%%%%%%%%%%%%%%%%%%%%%%%%%%%%%%%%%%%%%%%%%%%%%%
\section{Computational Modeling Results}
\subsection{Computational analysis}\label{sec-SimEandSig}
We investigated the tensile elastic modulus and strength assuming perfectly distributed platelets first. In this condition, no spatial variation was considered and $A_{11,avg}$ was set to be $0.5$. We extended the investigated thickness up to $91$ layers through the thickness ($12.7$ mm) to make sure to fully explore the thickness effect. Since our model is stochastic, we simulated $30$ coupons per case in order to get a good representation of the statistical behavior. The results are shown in Fig.~\ref{f-modulus} and Fig.~\ref{f-strength} where it can be noted that the model matches the experimental average tensile modulus and strength. Thanks to the computational analysis, we find that the modulus and strength reach an asymptotic value after an average number of layers of $45$ ($6.35$ mm). At $91$ layers ($12.7$ mm) thickness, the normalized modulus of both platelet morphologies reaches $0.97$. For the strength, they reach $0.49$ for the narrow and $0.45$ for the square (only $7\%$ difference). 

To summarize, the platelet width effect diminishes in both modulus and strength after DFC coupons reach the asymptotic thickness of $45$ layers ($6.35$ mm). Therefore, an important takeaway from the analysis is that for sufficiently large average numbers of layers through the thickness the effects of the platelet morphology and structure thickness become negligible and do not need to be accounted for design. On the other hand, for numbers of platelets through the thickness lower than about $45$, the morphology and structure thickness effects become dominant. Engineers and designers working in such a structure thickness range must take particular care and account for the expected reduction in modulus and strength.

Similar to the experiments, the simulated modulus and strength CoVs follow a decreasing trend as the thickness increases. However, the magnitude of the CoVs is significantly underpredicted in all the thicknesses and platelet sizes. For the $27$-layer case ($3.81$ mm), the narrow platelet coupons have a modulus CoV of $12\%$ from the experiments but only $1.2\%$ from the simulations. The coupons made using square platelets have an experimental CoV of $7\%$ while $3.2\%$ is predicted by the simulations. Such discrepancies are due to the assumption of an ideal platelet orientation throughout the specimen while, in reality, we know that the orientation tensor can vary spatially within the specimen, leading to higher CoVs than those predicted by the ideal case. To better understand the link between the spatial distribution of the orientation tensor and the CoVs on mechanical properties, we closely investigate the effect of the platelet orientation variations in the following section.

%%%%%%%%%%%%%%%%%%%%%%%%%%%%%%%%%%%%%%%%%%%%%%%%%%%%%%%%%%%%
\subsubsection{Effect of the platelet orientation variations}
To find the platelet orientation variations, we first find the average platelet orientation ($A_{11,avg}$) within $25.4~\text{mm} \times 25.4~\text{mm}$ SRVEs. Then we define two orientation variations. Global orientation variation finds the CoV of $A_{11,avg}$ from the entire SRVE population. If there are $10$ coupons with $6$ SRVEs each, the entire SRVE population size is $60$. Local orientation variation measures the CoV of $A_{11,avg}$ from SRVEs within a coupon. Using the global and local orientation variation, we can easily manipulate the statistical variation within and in between DFC coupons.

\begin{figure}
 \centering
\includegraphics[trim=0cm 3cm 0cm 0cm, clip=true,width = 1\textwidth, scale=1]{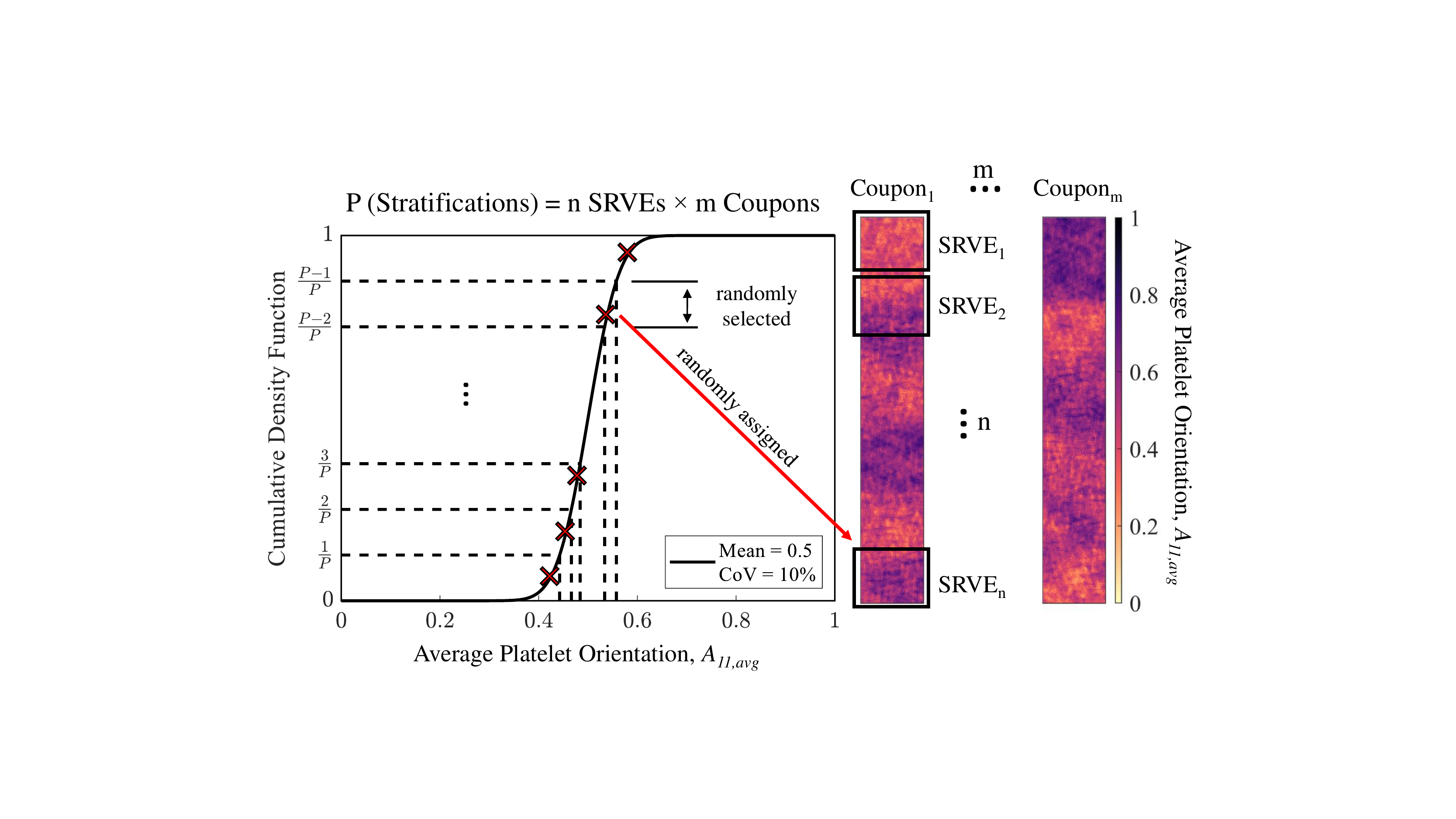} \caption{\label{f-LHS} Description of the Latin Hypercube Sampling (LHS) method to create coupons with random $A_{11,avg}$s. The Gaussian distribution is used to describe the distribution of $A_{11,avg}$s. LHS effectively assigns the global platelet orientation variation following the given mean and variation.} 
\end{figure}

Without the spatial variation, the global orientation variation is $\sim 2\%$ for the narrow platelets and $\sim 5\%$ for the square platelets at the thickness of 27 layers ($3.81$ mm). From the $\micro$CT scan, the global orientation variation of the square platelet coupons is equal to $9.2\%$. As can be noted, a subtle difference exist between the simulation and the experiment. We perform a parametric study by controlling the global orientation variation. We set the mean $A_{11,avg}$ at $0.5$ and increase the CoV by $2, 10,$ and $15\%$. To efficiently assign $A_{11,avg}$ into SRVEs, we use the Latin-Hypercube Sampling (LHS) method \cite{McKay1980}. The LHS is chosen because it guarantees to cover the two end-tails of the probability distribution. We assume that the the average platelet orientation, $A_{11,avg}$, follows the Gaussian distribution.  The assumption is based on the central limit theorem which states that the distribution of the sample averages tends to the Gaussian distribution as the independent random sample population increases \cite{Walpole1993}. Since the $A_{11,avg}$ is the average of independent random platelet orientations, the distribution of the entire $A_{11,avg}$ within a coupon or a plate becomes the normal distribution.  We calculate the cumulative density function (CDF) from the selected mean and CoV. The CDF is divided into equal cumulative probability spaces, called stratifications. The total number of stratifications equals to a number of SRVEs per coupon ($6$ of $1" \times 1"$ SRVE) times a total number of coupons. We simulate $30$ coupons per CoV, ending up with $180$ stratifications. Within a stratification, a random point is selected. This random point corresponds to a specific CDF value and therefore unique $A_{11,avg}$. Finally, we assign $6$ random $A_{11,avg}$ into a coupon until all coupons have assigned $A_{11,avg}$. In Fig.~\ref{f-LHS}, we illustrate the process of LHS method. Using the LHS, we can easily control the global orientation variation of the entire coupon population.

\begin{figure}
 \centering
\includegraphics[trim=0cm 3cm 0cm 0cm, clip=true,width = 1\textwidth, scale=1]{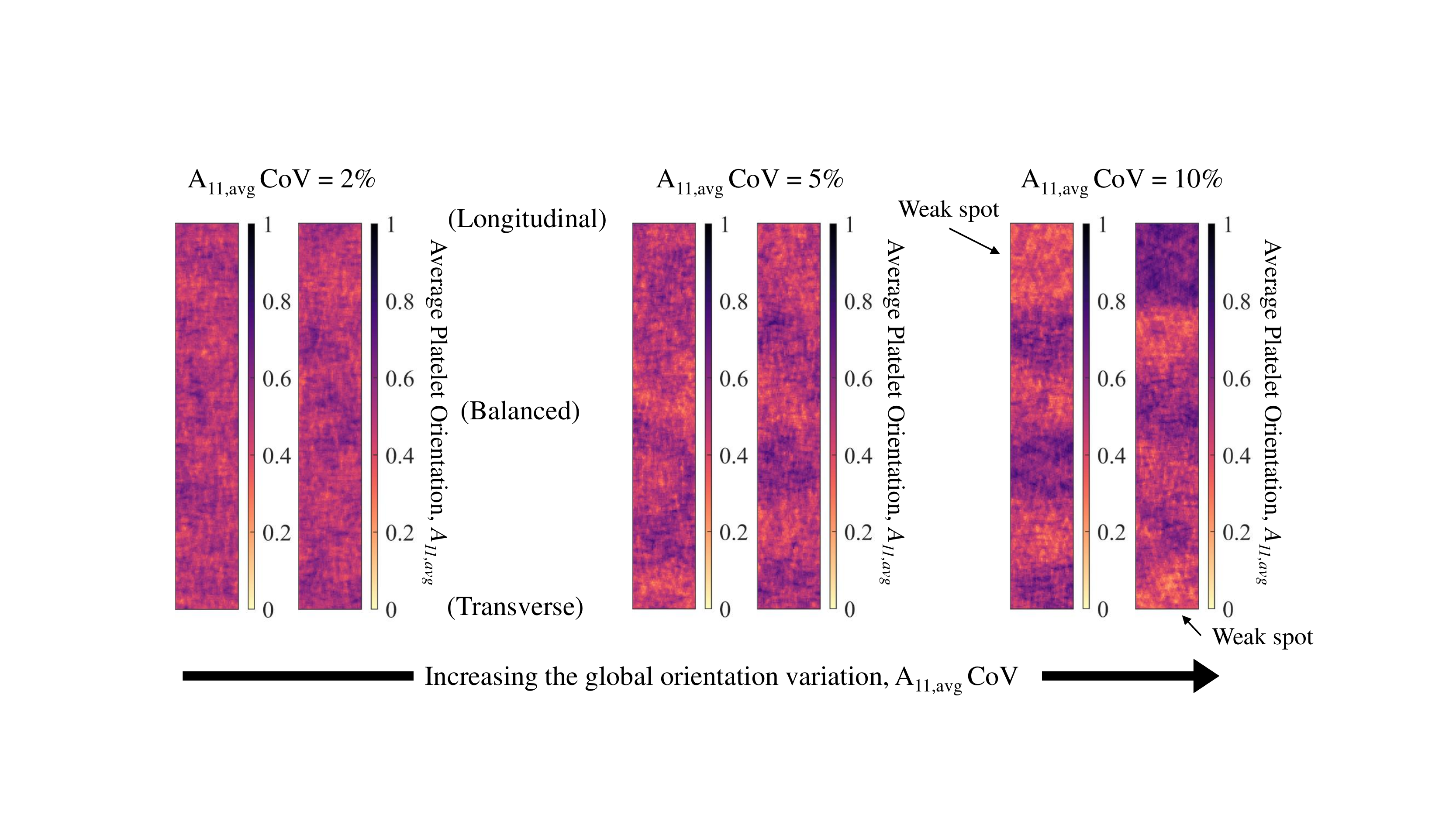} \caption{\label{f-LHSCoupons} Visualization of the platelet orientation distribution with different global orientation variations. When the global variation increases, it shows that the DFC coupons have noticeable regions with poorly oriented platelets (local weak spots). The average coupon orientations remain relatively constant at $0.5$.} 
\end{figure}

\begin{figure}
 \centering
\includegraphics[trim=0cm 3cm 0cm 0cm, clip=true,width = 1\textwidth, scale=1]{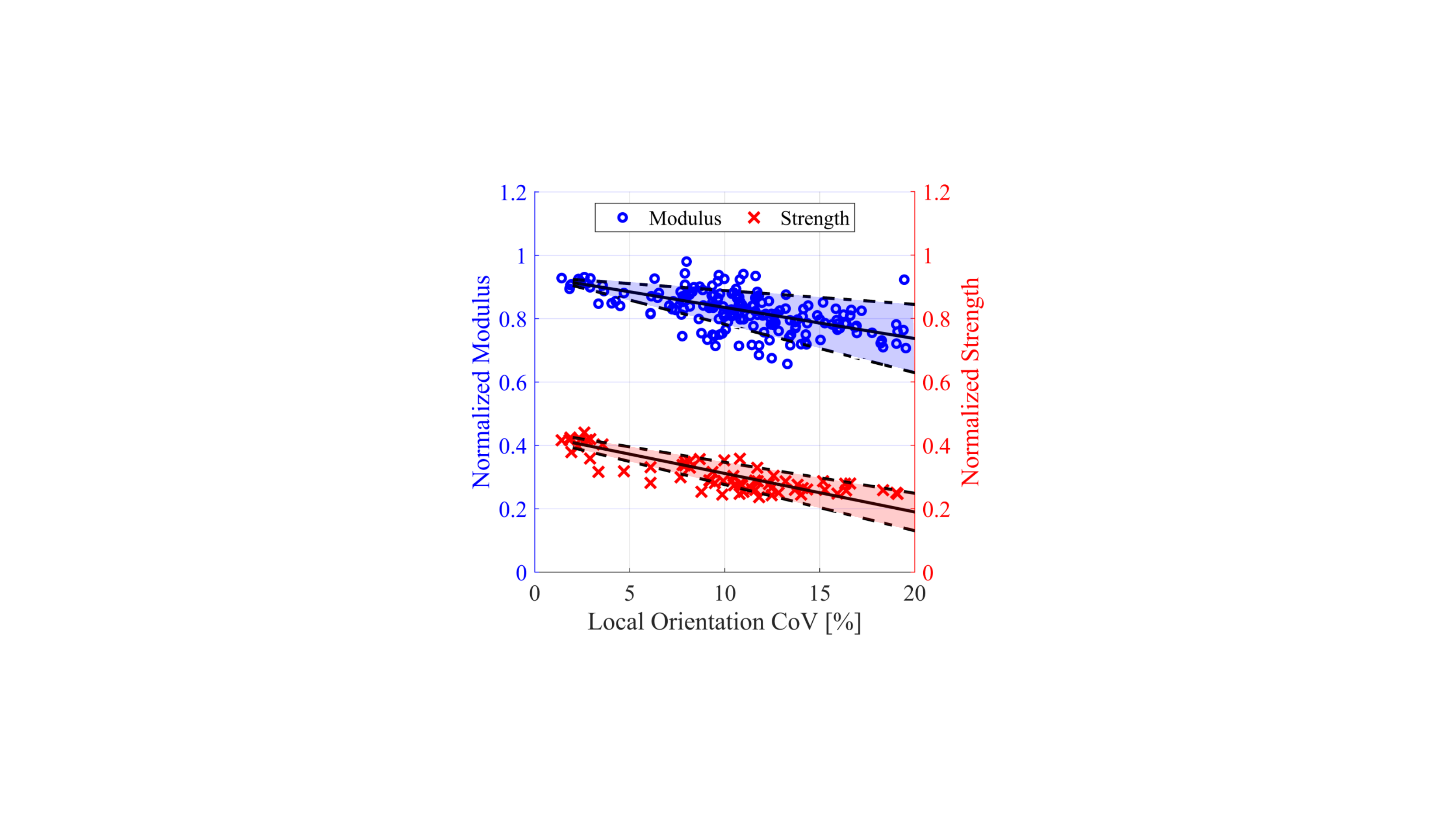} \caption{\label{f-A11CoVResult} Simulated normalized modulus and strength against the local platelet orientation variation. Increasing the local platelet orientation variation decreases the average modulus and strength (solid line) while increasing their variations (dashed lines). The narrow platelets with coupon thicknesses of $3.81$ mm ($27$ layers) are simulated. The square platelets also possess similar trend.} 
\end{figure}

Increasing the global orientation variation directly affects the local orientation distributions of the platelets. In Fig.~\ref{f-LHSCoupons}, we increase the global orientation CoV from $2\%$ to $10\%$. Simulated coupons are made with narrow platelets with $27$ layers ($3.81$ mm) through the thickness. We can clearly notice $A_{11,avg}$ varies within a coupon as the global orientation CoV increases. Increasing the orientation variation induces regions with poorly oriented platelets which can trigger fracture at lower strength. We plot the normalized modulus and strength against the local orientation variation in Fig.~\ref{f-A11CoVResult} where all the investigated coupons have a thickness of $27$ layers ($3.81$ mm) with narrow platelets. Solid lines represent the averages and dashed lines represent one standard deviation. Clearly, the strength and the modulus decrease as the local orientation CoV increases. The average strength is reduced more than the average modulus because the strength is more sensitive to the presence of local weak spots \cite{Weibull1951, Sal21a, Baz2017, Baz2004}. 

\begin{figure}
 \centering
\includegraphics[trim=0cm 3cm 0cm 0cm, clip=true,width = 1\textwidth, scale=1]{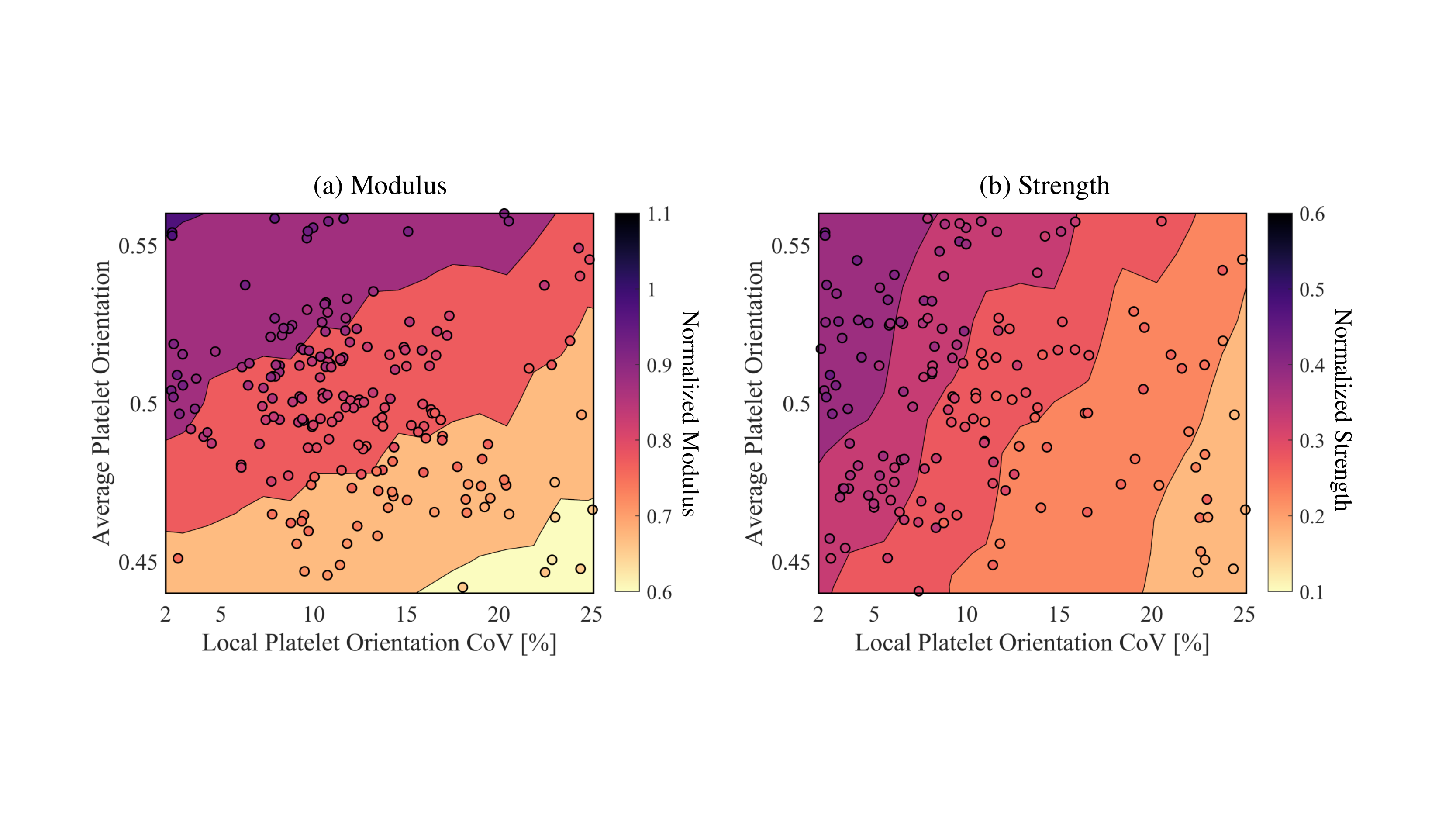} \caption{\label{f-EandSigContourPlot} Simulated (a) normalized modulus and (b) normalized strength against the local platelet orientation variation and the average platelet orientation. Comparing the slopes of contour lines in (a) and (b), we can determine the dominating parameter for the modulus and strength. (a) For the modulus, the average platelet orientation is the dominating parameter. However, the variation also plays a significant role where it can significantly reduce the modulus. (b) For the strength, the local platelet orientation variation is the dominating parameter.} 
\end{figure}

From Fig.~\ref{f-A11CoVResult}, we observe the effect of the platelet orientation variation on the mechanical properties of DFCs. Besides the orientation variation, there is another important mesostructure characteristic that defines DFC structures, which is the average platelet orientation. The average platelet orientation determines the coupon modulus according to many studies \cite{Denos2018a,Li2017,Chen2018,Wan2016a,Favaloro2018,Kravchenko2017,Martulli2019,Fera2010,Sel2017,Ko2021,Shah2020,Wan2021}. We combine two different mesostructure characteristics and show the modulus and strength change in Fig.~\ref{f-EandSigContourPlot} where we plot contour maps of the modulus and strength in using a Delaunay triangulation to perform the interpolation between the data ~\cite{Amidror2002}. Comparing between Fig.~\ref{f-EandSigContourPlot}a and b, the contour lines in the modulus are tilted more horizontally than the strength. For the modulus, the average platelet orientation is the dominating parameter. However, the effect of the orientation variation is not negligible. We can explain this relationship with a series coupling of individual SRVEs. As described in Eq.~\ref{eqModulus}, each modulus in SRVE contributes to the equivalent modulus \cite{Mindess2003}.
\begin{equation}\label{eqModulus}
E_{equivalent} = \left (\sum_{i=1}^{N}\frac{1}{E_i}\right )^{-1}
\end{equation}
Because this equation is composed of the product and summation of each modulus in SRVE, both average and variation of the modulus are important. The modulus and the platelet orientation, $A_{11,avg}$ are also directly related. Therefore, to accurately predict the modulus, we need both average platelet orientation and their CoVs. 

The platelet orientation CoV plays more significant role in case of the strength than the modulus (see Fig.~\ref{f-EandSigContourPlot}b). Compared to the modulus, the strength contour lines are tilted more vertically. Unlike the modulus, the strength is more dependent on the individual SRVE's probability of survival. If a single SRVE fails, the whole structure fails \cite{Weibull1951,Baz2004,Baz2017,Sal21a}. It can be simply described as a finite weakest link chain model described in following equation where $P_{f}(\sigma_N)$ is the failure probability for a given nominal stress: 
\begin{equation}\label{eqStrength}
P_f(\sigma_N)=1-\prod_{i=1}^{N}(1-P_{fi}(\sigma_N))
\end{equation}
The weakest SRVE or the SRVE having the most poorly oriented platelets governs the strength of the coupon. Therefore, the average platelet orientation alone is limited to indicate the strength of DFC structures.

%%%%%%%%%%%%%%%%%%%%%%%%%%%%%%%%%%%%%%%%%%%%%%%%%%%%%%%%%%%%
\subsubsection{Matching the modulus and strength variations}
\begin{figure}
 \centering
  \includegraphics[trim=0cm 1cm 0cm 1cm, clip=true,width = 1\textwidth, scale=1]{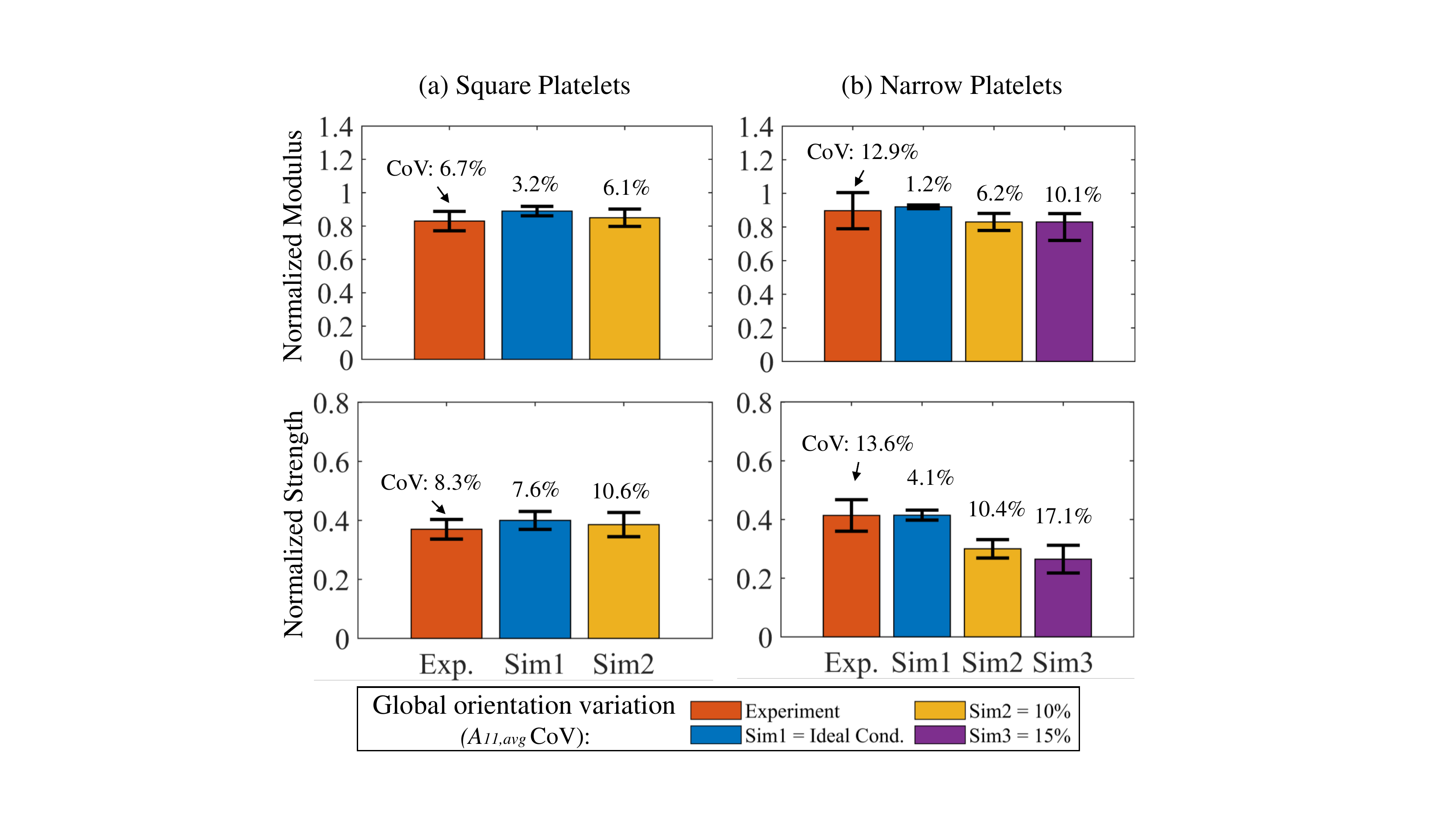} \caption{\label{f-COVParametricStudy} Normalized modulus and strength with meso-structures made of different platelet orientation variation. For (a) the square platelets, the meso-structure with input $A_{11,avg}$ CoV of $10\%$ precisely match the modulus and strength variation from the experiment. However, for (b) the narrow platelets, CoV of $10\%$ still underestimates the variation from the experiment. In fact, it is difficult to match both modulus and strength CoV using a single input CoV. In both platelet sizes, the ideal manufacturing condition underpredicts the global orientation variation.}
\end{figure}

By manipulating the global orientation variation, we can match the modulus and strength variation from the experiment. We take advantage of the spatial variation algorithm to calibrate the mesostructures with different global orientation variations. Both platelet sizes are simulated with coupon thickness of $27$ layers ($3.81$ mm). At least $30$ coupons are simulated per global orientation variation. The results are plotted in 
Fig.~\ref{f-COVParametricStudy}. By manipulating the global orientation variation, we can control the modulus and strength variations. 

In case of the square platelet, we successfully match both average and variation associated with the tensile modulus and strength (see Fig.~\ref{f-COVParametricStudy}a). We use the average platelet orientation at $0.5$ and the global orientation variation of $10\%$ as the calibration parameters. The global orientation variation of $10\%$ agrees with the $\micro$CT measurement of $9.2\%$. These results indicate that if we know the statistical information of the platelet orientation prior to the experiment, we can predict not only the average but also the associated variation of both modulus and strength of DFC coupons. It will significantly reduce the required number of proof testing in order to certify these stochastic material properties.

In case of the narrow platelets, we reach the experimental variation using the $15\%$ orientation CoV. Without calibrating the spatial distribution of the platelets, such variations are hard to obtain. However, we observe noticeable reduction of the average strength as the orientation CoV increases. We believe this mismatch is due to the fact that we assume the size of the SRVE is $25.4~\text{mm} \times 25.4~\text{mm} $. Changing the size of the SRVE will affect the average and variation of the strength. We will investigate such effect in future studies.

%To validate the computational results, we use the X-ray $\micro$CT scans to find the statistical information of the platelet orientations. We compare the global orientation variation used it to calibrate the meso-structure against the measurement directly from the CT scans. - Don't need this paragraph anymore. SK

\begin{figure}
 \centering
  \includegraphics[trim=0cm 2cm 0cm 2cm, clip=true,width = 1\textwidth, scale=1]{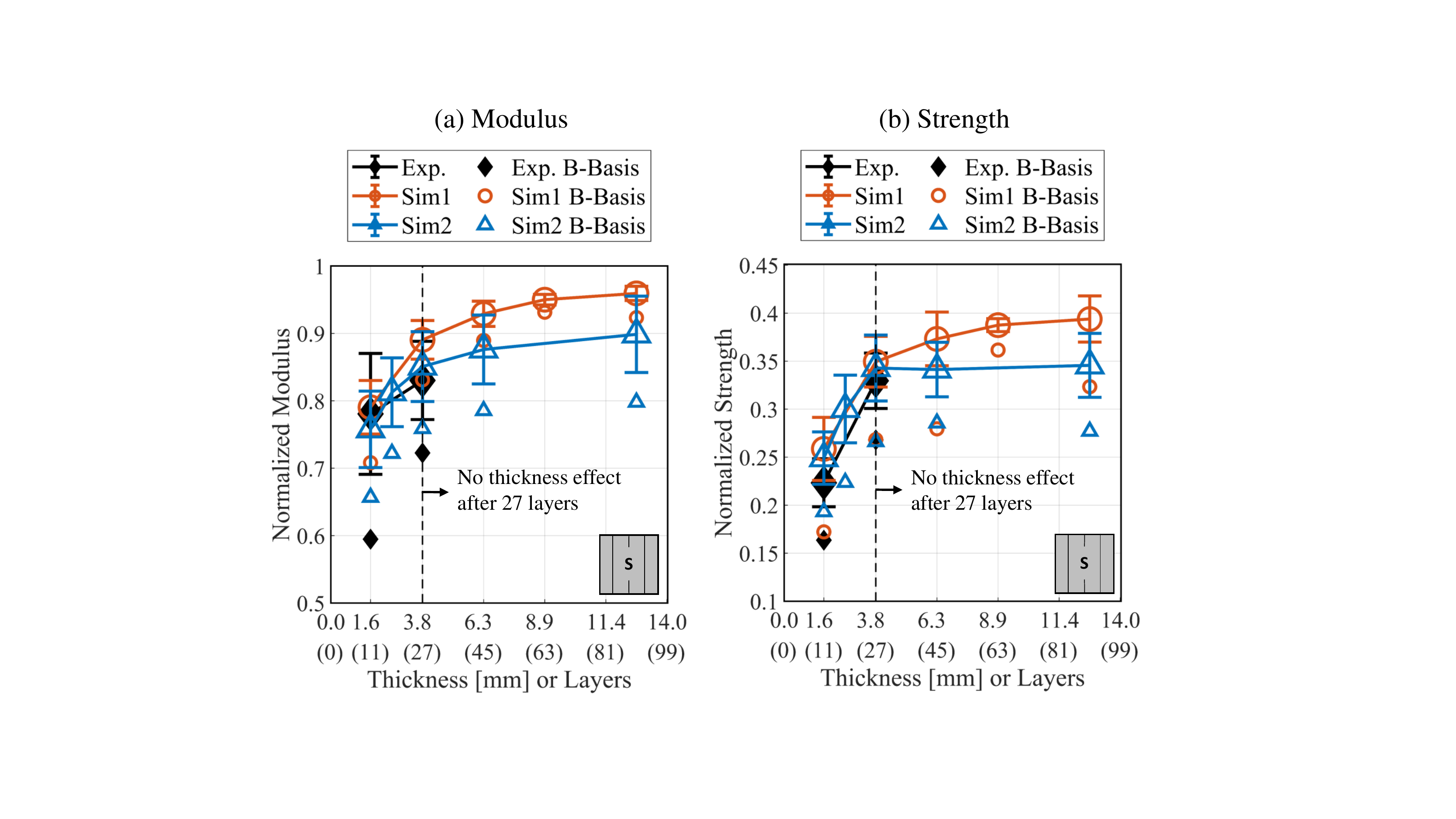} \caption{\label{f-EandSigWithNewCoV} The square platelet thickness effect compared between the experiment and simulations with different meso-structures. Sim1 represents the ideal manufacturing condition where $A_{11,avg}\text{CoV} = 5\%$. Sim2 represents the calibrated meso-structure using the $\micro$CT scanning where $A_{11,avg}\text{CoV} = 10\%$. Using the calibrated meso-structures, we found the thickness asymptotic level moved from $45$ layers to $28$ layers in both (a) modulus and (b) strength. The thickness effect is influenced by both the average platelet orientation and the global orientation variation within the coupons.}
\end{figure}

With the calibrated mesostructures, we simulated different coupon thicknesses of the square platelet coupons. The results are shown in Fig.~\ref{f-EandSigWithNewCoV}. The simulation results with the calibrated mesostructures (marked in $A_{11,avg} \text{CoV} = 10\%$) follow the experimental results more precisely than the ideal manufacturing condition simulation results (shown in $A_{11,avg} \text{CoV} = 5\%$). At the thickness of $91$ layers ($12.7$ mm), the average strength is $0.391$ with CoV of $9.7\%$. The B-basis value is $0.314$ showing $15.3\%$ difference compared to the non-calibrated meso-structure simulations. More importantly, the asymptotic thickness level moved from $45$ layers to $28$ layers ($\sim 46\%$ difference). This clearly shows that, to accurately capture the thickness effect, it is important to capture both the average orientation tensor of the platelets and also their associated global orientation variation.

%%%%%%%%%%%%%%%%%%%%%%%%%%%%%%%%%%%%%%%%%%%%%%%%%%%%%%%%%%%%
\subsubsection{Effect of the resin-rich volume fraction}
Besides the platelet orientation, we also investigate the effect of the resin-rich volume fraction. Large amount of resin-rich volumes are an unavoidable mesostructure characteristic of DFCs due to the discontinuity of the platelets. Volume fractions of $5 \sim 10\%$ of the total volume \cite{Sel2016, Sommer2020b} are not uncommon but are often overlooked. This is because the presence of platelets in multiple directions that can carry load makes the effect of porosity in DFCs less severe than in traditional composites. 

Wan and Takahashi \cite{Wan2016a} experimentally showed that the tensile and compressive strength changed with respect to the fiber volume fraction. They measured the fiber volume fraction with respect to the molding pressures at $5$ and $10$ MPa. When the pressure was reduced, the fiber volume fraction also decreased from $\sim56\%$ to $\sim50\%$. The decreased fiber volume fraction aligned with the reduction of the tensile and compressive strength. We further investigate the effect of resin-rich volume fraction leveraging the comprehensive random mesostructure generation algorithm.

From the algorithm, we indirectly control the total percentage of the resin-rich layers. In the generation algorithm, we limit the number of maximum platelets that can be deposited in the partitions. Currently, the maximum limit is set as one standard deviation (or CoV~$= 22\%$) above the target average. This value is experimentally determined from previous studies \cite{Ko2019a, Ko2019b}. By changing the maximum limit of the platelets, we can vary the percentage of the resin-rich layers. We varied the resin-rich percentage from $4$ to $8\%$. The simulated coupons had the narrow platelets, thickness of $27$ layers ($3.81$ mm), and no spatial variation to explicitly observe the effect of the resin-rich layers only.

\begin{figure}
 \centering
\includegraphics[trim=0cm 4cm 0cm 0cm, clip=true,width = 1\textwidth, scale=1]{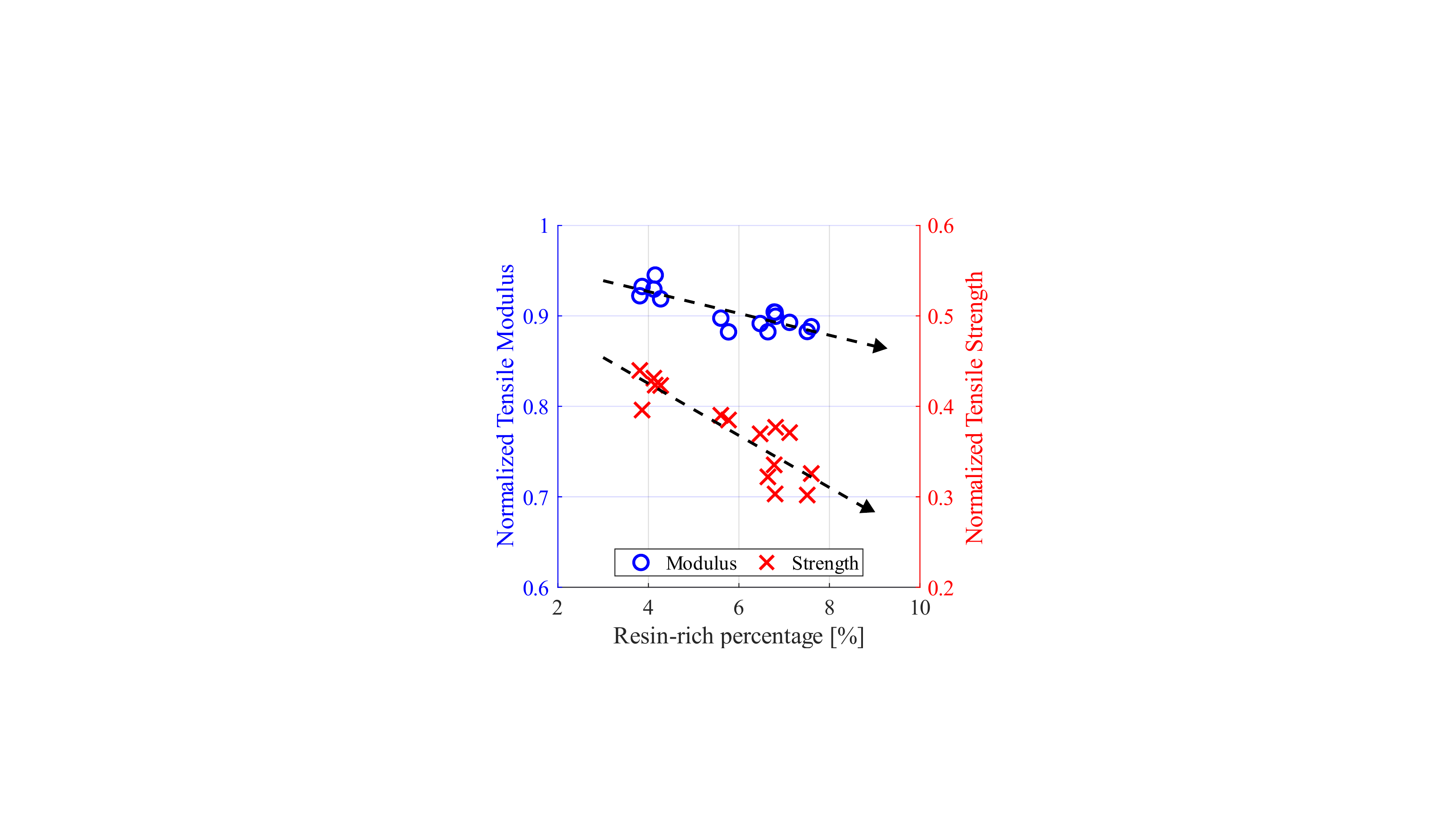} \caption{\label{f-ResinRich} Simulated normalized modulus and strength against varying resin-rich percentage of an entire coupon volume. The resin-rich percentage significantly penalizes the strength of DFC coupons. The narrow platelets with coupon thicknesses of $28$ layers are simulated.} 
\end{figure}

In Fig.~\ref{f-ResinRich}, the average modulus and strength with respect to the resin-rich percentage are plotted. The figure shows a clear linear decreasing trend of the modulus and strength as the resin-rich percentage increases. The dashed line represents a linear fit of the modulus and strength. As can be noted, a small percent change, from $4$ to $8\%$, of the resin-rich layers causes a significant reduction ($25\%$) of the strength. The resin-rich layers have direct impact to the mechanical performances of DFCs. To minimize the effect of the resin-rich layers, we recommend to precisely optimize the molding pressure. 

%%%%%%%%%%%%%%%%%%%%%%%%%%%%%%%%%%%%%%%%%%%%%%%%%%%%%%%%%%%% 
\subsection{Platelet width effects}\label{sec-PlateletWidthEffects}
From the experiment, we observed significant platelet width effect on the tensile modulus and strength (see Figs.~\ref{f-modulus} and \ref{f-strength}). As the platelet width increases, both the tensile modulus and strength decrease. In this section, we discuss the mesostructure characteristic that governs the platelet width effect.

\begin{figure}
 \centering
\includegraphics[trim=0cm 4cm 0cm 0cm, clip=true,width = 1\textwidth, scale=1]{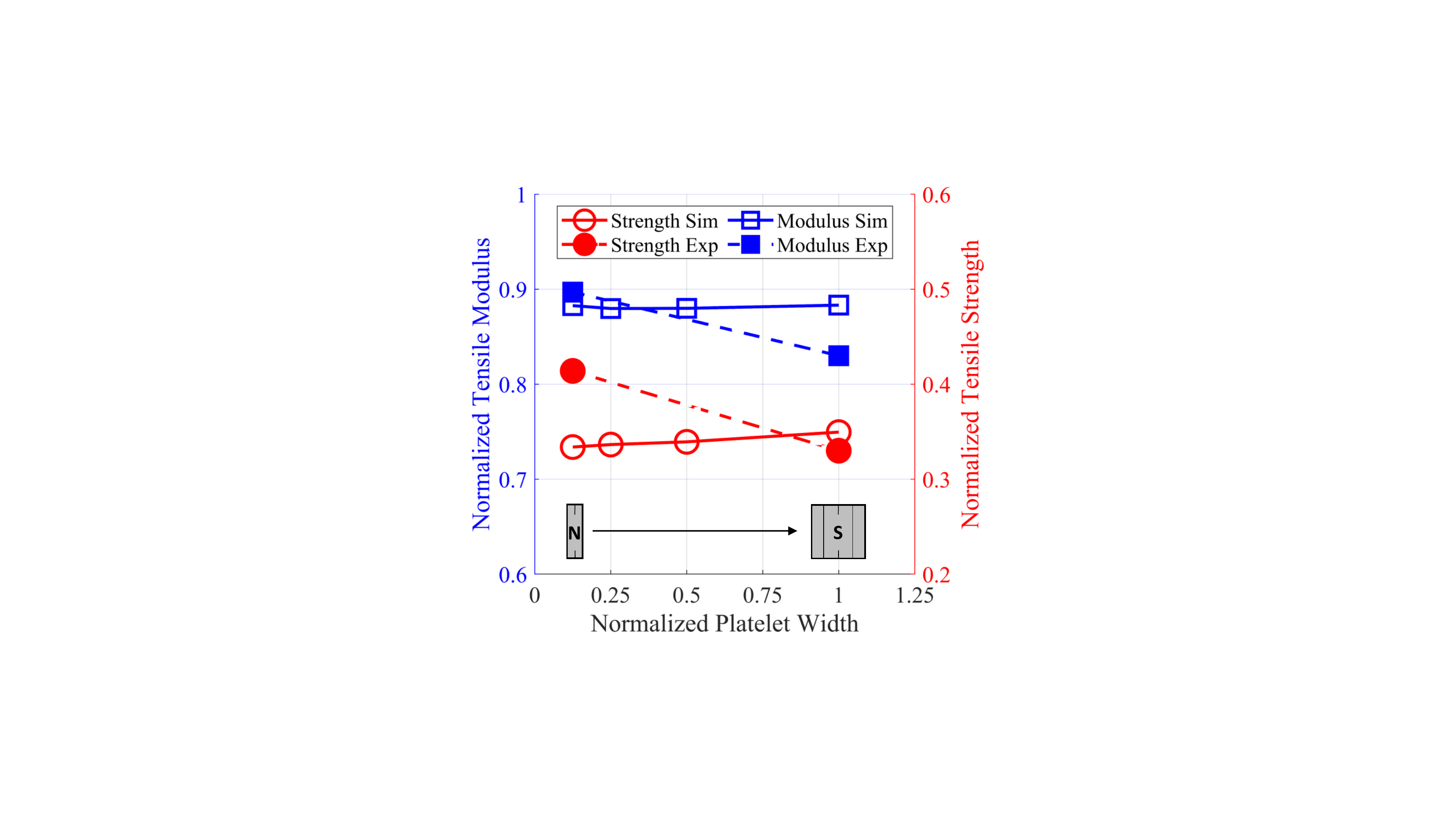} \caption{\label{f-PlateletWidthEffect} The effect of the global orientation variation in DFCs. Here, we simulated different platelet widths with fixed global orientation variation of $5\%$, the $A_{11,avg}$ of $0.5$, and the coupon thickness of $27$ layers ($3.81$ mm). As clearly shown in the experiment, the platelet size affects the modulus and the strength. However, if the global orientation variation is fixed, the platelet width effect vanishes. Therefore, we can confirm that the global orientation variation is the root cause of the platelet width effect in DFCs.}

%(a) As the platelet width increases, we observe reduction in the modulus and strength. The strength reduction is especially noticeable. (b) As the platelet width increases, the meso-structures introduce higher local platelet orientation variations. Higher variation creates local weak spots. Therefore, the coupon strength reduces as the platelet width increases. THIS MUST BE CHANGED} 
\end{figure}

As we discussed in the section \ref{Sec-ExperimentalResults}, the platelet width controls the global orientation variation. Therefore, the modulus and strength variations with respect to the platelet width must be closely related to the global orientation variation. In other words, if the platelet width is changed but the global orientation variation is fixed, then the modulus and strength should not be changed. By leveraging the stochastic modeling, we simulated four different platelet widths with the constant global orientation variation of $\sim5\%$. The $A_{11,avg}$ was $0.5$ and the coupon thickness was $27$ layers ($3.81$ mm). We simulated $30$ coupons per platelet width. In Fig.~\ref{f-PlateletWidthEffect}, the tensile stiffness and strength of different platelet widths are plotted. Also, we plot the experimental results for the comparison. When the global orientation variation is fixed, the tensile modulus and strength are relatively unchanged even though the platelet width are significantly changed. However, these results are opposite to what we have observed from the experiments. The only difference is that the global orientation variation is fixed when the platelet width is altered. These results confirm that the platelet width effect is originated from the global orientation variation.

%%% This was the original paragraph 8/2/2022
%From the experiment (see Figs.~\ref{f-modulus} and \ref{f-strength}), we observe a significant reduction in the tensile strength as the platelet width increases. In this section, we discuss what are the mesostructure characteristics that drive such platelet width effect. We simulated four different platelet widths at the coupon thickness of $27$ layers ($3.81$ mm). We simulated $10$ coupons per platelet width. The coupons were generated using the ideal manufacturing condition. In Fig.~\ref{f-PlateletWidthEffect}a, the tensile stiffness and strength of different platelet widths are plotted. As the platelet width increases by $\sim8$ times, the tensile strength is decreased by $17\%$. The stiffness is much less affected. We observe three mesostructure characteristics. They are the average platelet orientation, platelet orientation variation, and resin-rich volume fraction. As the platelet width changes, the average platelet orientation and the resin-rich percentage remain relatively constant but the orientation variation significantly changes. As can be seen in Fig.~\ref{f-PlateletWidthEffect}b, the orientation variation is increased more than double. Increasing the orientation variation reduces the strength by introducing the local weak spots (see Fig.~\ref{f-LHSCoupons} and \ref{f-EandSigContourPlot}). Therefore, the narrow platelets are recommended not only because they have the manufacturing advantage, but they also possess higher tensile strength.

%%%%%%%%%%%%%%%%%%%%%%%%%%%%%%%%%%%%%%%%%%%%%%%%%%%%%%%%%%%%
\subsection{Structure thickness effects}
\begin{figure}
 \centering
\includegraphics[trim=0cm 3cm 0cm 0cm, clip=true,width = 1\textwidth, scale=1]{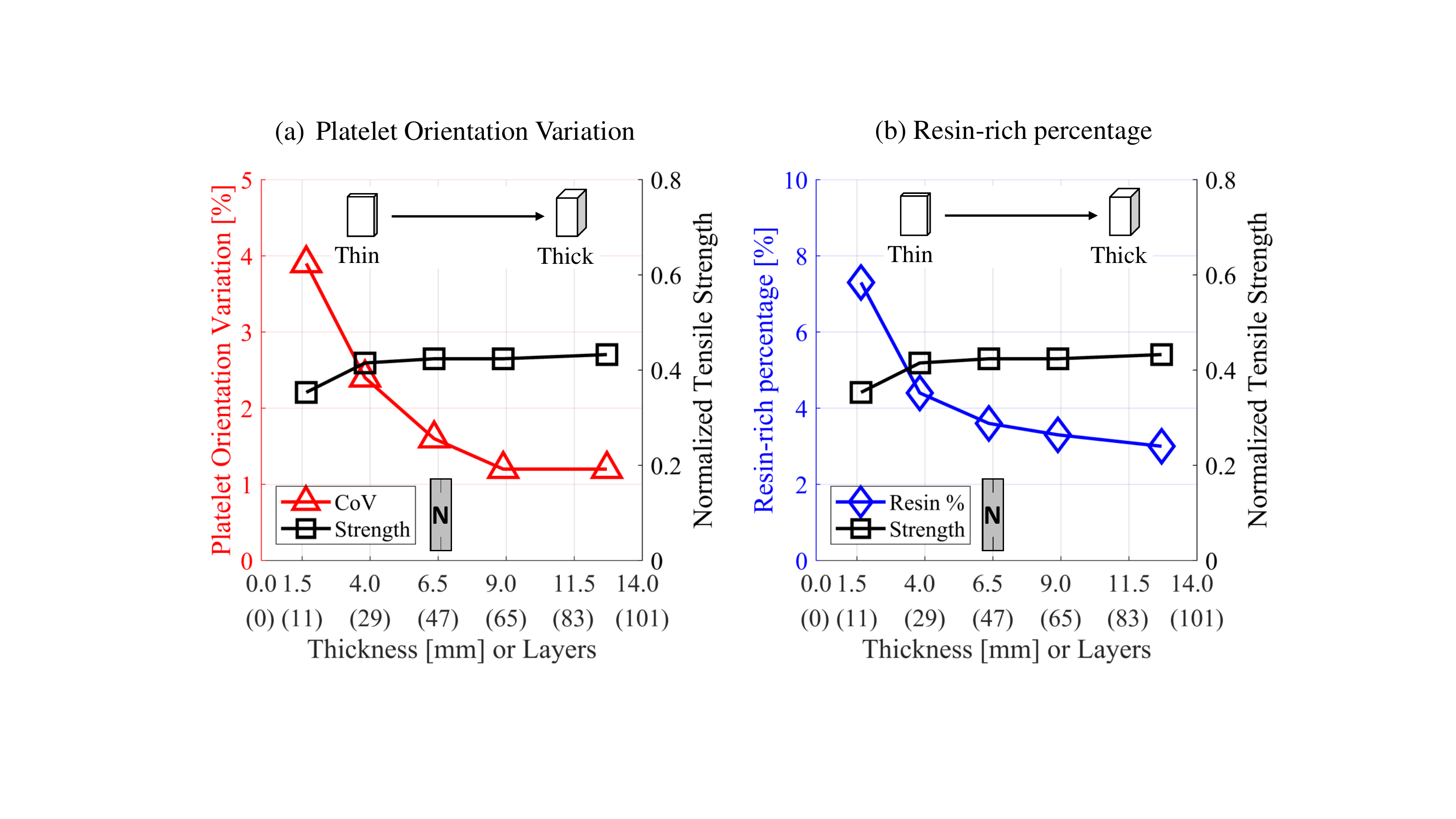} \caption{\label{f-ThicknessEffect} As the coupon thickness changes, we observe two meso-structure characteristics change. (a) The platelet orientation variation decreases and (b) resin-rich percentage also decreases as the coupon thickness increases. Increasing the coupon thickness significantly benefits the strength because it reduces having a chance of local weak spots.} 
\end{figure}

Kravchenko \textit{et al}. \cite{kravchenko2019a,kravchenko2019b} pointed out that the thickness effect was due to the change of uniformity of the local orientation state. Similar to the platelet width effect, we investigate the three mesostructure characteristics with respect to the coupon thickness. We simulated $30$ coupons made of the narrow platelets. The simulated meso-structures based on the ideal manufacturing condition. The platelet orientation variation and the resin-rich percentage are strongly affected by the coupon thickness while the the average platelet orientation remains close to $0.5$ in all cases. In Fig.~\ref{f-ThicknessEffect}, we plot the results. As the coupon thickness increases, the tensile strength increases until the thickness of $45$ layers ($6.35$ mm). We clearly see that both orientation variation and the resin-rich percentage decrease as the the coupon thickness increases. Especially, the resin-rich layers decrease drastically as the thickness increases. By increasing the coupon thickness, the average number of platelets through the thickness increases. This reduces the probability of having poorly oriented regions as well as the resin-rich layers which can trigger fracture.

\section{Conclusions}
In this study, we investigate the tensile modulus and strength of Discontinuous Fiber Composites (DFCs) as a function of the average number of platelets through the thickness of the structure and the platelet width. We analyze the mechanical properties using experiments and stochastic finite element modeling. From the results, we make the following conclusions:

\be  \setlength{\itemsep}{1mm}
\ii DFCs exhibit strong thickness effect on the tensile modulus and strength. From the experiments, we observe a $1.4\%$ and $22.4\%$ increase in the modulus and strength of coupons made with narrow platelets when the thickness increases from $11$ layers ($1.65$ mm) to $45$ layers ($6.35$ mm). For the case of coupons made of square platelet, we observe $6.4\%$ and $47.8\%$ increase in the modulus and the strength when the thickness increases from $11$ layers ($1.65$ mm) to $27$ layers ($3.81$ mm). 

\ii Samples made with narrow platelets outperform the ones made of square platelets in both average modulus and strength. At the thickness of $27$ layers ($3.81$ mm), the narrow platelet outperforms the square by $8.0\%$ in modulus and $25.4\%$ in strength. However, the coupons with narrow platelets also show significantly higher Coefficient of Variations (CoVs) than those with square platelets ($12.9\%$ vs $6.7\%$ in the modulus and $13.6\%$ vs $8.3\%$ in the strength). Therefore, a trade off exists between the average and variation of the mechanical properties when the platelet width changes. 

\ii Using stochastic finite element models, we capture the thickness effect in both the modulus and the strength. We extend the coupon thickness up to $91$ layers ($12.7$ mm). For both platelet types asymptotic modulus and strength are reached for thickness of $45$ layers ($6.35$ mm) or higher in case of perfectly distributed platelets. At the asymptotic limit, the platelet width effect and the structure thickness effect become negligible. In fact, the asymptotic modulus reaches $96\%$ of the theoretical modulus of a quasi-isotropic (QI) laminate. For the strength, the narrow and the square platelet reach $43\%$ and $39\%$ of the QI laminate strength.

\ii Leveraging the computational modeling equipped with the spatial variation of the platelets, we study the effect of the local orientation variation within coupons. For both the modulus and the strength of DFC structures, the average platelet orientation and the local platelet orientation variation play a significant role. For the modulus, the average platelet orientation plays the dominant role while the local orientation variation plays the key role for the strength. Therefore, we must consider both characteristics when relating the DFC meso-structures to their tensile material properties.

\ii An additional important mesostructure characteristic that defines the DFC material properties is the resin-rich volume fraction. The thickness effect is caused by both the change in the local platelet orientation variation and the resin-rich percentage. The two characteristics significantly change as the thickness varies up to the asymptotic thickness limit of $45$ layers ($6.35$ mm). The platelet width effect is mainly caused by the change in the local platelet orientation variation.

\ii Lastly, we suggest two alternative mesostructure calibration methods to accurately capture both average and variation of the material properties. First is to calibrate the mesostructures against the experimental results. Using the Latin-hypercube sampling method, we find the accurate platelet orientation variation. Second is to use the X-ray $\micro$CT scans to directly obtain the platelet orientation distributions. Two independent methods will provide accurate estimation of the B-basis design values for engineers and designers.

\ii In future work, we will investigate how to find the SRVE size corresponding to the platelet geometry. The SRVE size should be determined based on the material's unique inhomogeneity characteristic length. According to our previous studies \cite{Ko2019a, Ko2019b}, the effective length of the Fracture Process Zone (FPZ) changes with the platelet sizes. Therefore, the SRVE size depends on the platelet size as well. When the SRVE size is determined, we can further improve the accuracy of the stochastic finite element models.

\ee
 
 The foregoing results provide general guidance on the design and certification of DFC structures with varying thickness and platelet size. Also, they signify the importance of computational tools to minimize proof-testing required for the certification process.

%%%%%%%%%%%%%%%%%%%%%%%%%%%%%%%%%%%%%%%%%%%%%%%%%%
%%%%%%%%%%%%%%%%%%%%%%%%%%%%%%%%%%%%%%%%%%%%%%%%%%
%%%%%%%%%%%%%%%%%%%%%%%%%%%%%%%%%%%%%%%%%%%%%%%%%%
%%%%%%%%%%%%%%%%%%%%%%%%%%%%%%%%%%%%%%%%%%%%%%%%%%
\section*{CRediT author statement}
\textbf{Seunghyun Ko:} Conceptualization, Methodology, Software, Formal analysis, Investigation, Writing - Original Draft. \textbf{Troy Nakagawa:} Investigation, Resources, Writing - Review \& Editing. \textbf{Zhisong Chen:} Software, Investigation, Resources. \textbf{William B. Avery:} Validation, Project administration, Writing - Review \& Editing. \textbf{Ebonni J. Adams:} Validation, Resources, Project administration. \textbf{Matthew R. Soja:} Validation, Resources, Project administration. \textbf{Michael H. Larson:} Resources, Project administration. \textbf{Chul Y. Park:} Resources, Project administration. \textbf{Jinkyu Yang:} Resources, Project administration. \textbf{Marco Salviato:} Conceptualization, Methodology, Supervision, Project administration, Funding acquisition, Writing - Review \& Editing.

%%%%%%%%%%%%%%%%%%%%%%%%%%%%%%%%%%%%%%%%%%%%%%%%%%
%%%%%%%%%%%%%%%%%%%%%%%%%%%%%%%%%%%%%%%%%%%%%%%%%%
\section*{Acknowledgments}
We appreciate the financial support from the FAA-funded Center of Excellence for Advanced Materials in Transport Aircraft Structures (AMTAS) and the Boeing Company. We appreciate technical guidance and support from Dave Stanley, Ahmet Oztekin, Cindy Ashforth, and Larry Ilcewicz from the FAA, Scott Gunther from Boeing, Scott James and Robb Medved from Sekisui Aerospace, and Lisa Walton and Gary Bond from Solvay. We thank the Joint Center for Aerospace Technology Innovation (JCATI) and Chomarat for their project support. Lastly, we thank James Davey, Talal Abdullah, and Luke Kuklenski from the University of Washington for experimental help.


% \begin{appendices}
% %\appendix
% \section{Appendix A}
% \label{AppxA}

% Appendix A

% \section{Appendix B}
% \label{AppxB}
% \renewcommand{\theequation}{B.\arabic{equation}}
% \setcounter{equation}{0}

% Appendix B

% \end{appendices}

%%%%%%%%%%%%%%%%%%%%%%%%%%%%%%%%%%%%%%%%%%%%%%%%%%%%%%
%% Reference 
%%%%%%%%%%%%%%%%%%%%%%%%%%%%%%%%%%%%%%%%%%%%%%%%%%%%%%

\clearpage

\linespread{0}\selectfont
\bibliographystyle{unsrt}
\bibliography{main}

\begin{thebibliography}{10}

\bibitem{Barbero2018}
Barbero EJ.
\newblock {\em Introduction to Composite Materials Design}.
\newblock CNC Press, 2017.

\bibitem{Tsai2014}
D~Gay, SV~Hoa, and SW~Tsai.
\newblock {\em Composite Materials: Design and Applications}.
\newblock CRC Press, 2014.

\bibitem{Elmarakbi2013}
A~Elmarakbi.
\newblock {\em Advanced Composite Materials for Automotive Applications:
  Structural Integrity and Crashworthiness}.
\newblock Wiley, 2013.

\bibitem{Chortis2013}
DI~Chortis.
\newblock {\em Structural Analysis of Composite Wind Turbine Blades}.
\newblock Springer, 2013.

\bibitem{Boeing2021}
{Boeing 787 Dreamliner. https://www.boeing.com/commercial/787/; 2021}.

\bibitem{Baker2004}
A~Baker, S~Dutton, and D~Kelly.
\newblock {\em Composite Materials for Aircraft Structures}.
\newblock American Institute of Aeronautics and Astronautics, 2004.

\bibitem{Rashidi2018}
A~Rashidi and AS~Milani.
\newblock A multi-step biaxial bias extension test for wrinkling/de-wrinkling
  characterization of woven fabrics: Towards optimum forming design guidelines.
\newblock {\em Materials \& Design}, 146:273--285, 2018.

\bibitem{Wang2015}
P~Wang, X~Legrand, P~Boisse, N~Hamila, and D~Soulat.
\newblock Experimental and numerical analyses of manufacturing process of a
  composite square box part: Comparison between textile reinforcement forming
  and surface 3d weaving.
\newblock {\em Composites Part B: Engineering}, 78:26--34, 2015.

\bibitem{Sel2016}
M~Selezneva and L~Lessard.
\newblock Characterization of mechanical properties of randomly oriented strand
  thermoplastic composites.
\newblock {\em Journal of composite materials}, 50(20):2833--2851, 2016.

\bibitem{Denos2018a}
BR~Denos, DE~Sommer, AJ~Favaloro, RB~Pipes, and WB~Avery.
\newblock Fiber orientation measurement from mesoscale ct scans of prepreg
  platelet molded composites.
\newblock {\em Composites Part A: Applied Science and Manufacturing},
  114:241--249, 2018.

\bibitem{Jin2017}
BC~Jin, X~Li, A~Jain, C~Gonz{\'a}lez, J~LLorca, and S~Nutt.
\newblock Optimization of microstructures and mechanical properties of
  composite oriented strand board from reused prepreg.
\newblock {\em Composite Structures}, 174:389--398, 2017.

\bibitem{Li2017}
Y~Li, S~Pimenta, J~Singgih, S~Nothdurfter, and K~Schuffenhauer.
\newblock Experimental investigation of randomly-oriented tow-based
  discontinuous composites and their equivalent laminates.
\newblock {\em Composites Part A: Applied Science and Manufacturing},
  102:64--75, 2017.

\bibitem{Wan2016a}
Y~Wan and J~Takahashi.
\newblock Tensile and compressive properties of chopped carbon fiber tapes
  reinforced thermoplastics with different fiber lengths and molding pressures.
\newblock {\em Composites Part A: Applied Science and Manufacturing},
  87:271--281, 2016.

\bibitem{Chen2018}
Z~Chen, T~Huang, Y~Shao, Y~Li, H~Xu, K~Avery, D~Zeng, W~Chen, and X~Su.
\newblock Multiscale finite element modeling of sheet molding compound (smc)
  composite structure based on stochastic mesostructure reconstruction.
\newblock {\em Composite Structures}, 188:25--38, 2018.

\bibitem{Ryatt2022}
J~Ryatt and M~Ramulu.
\newblock Prediction of tensile failure of stochastic tow-based discontinuous
  composites via mesoscale finite element analysis.
\newblock {\em Composite Structures}, 279:114769, 2022.

\bibitem{Fatemi2015}
S~Mortazavian and A~Fatemi.
\newblock Fatigue behavior and modeling of short fiber reinforced polymer
  composites: A literature review.
\newblock {\em International Journal of Fatigue}, 70:297--321, 2015.

\bibitem{Boursier2001}
B~Boursier.
\newblock New possibilities with hexmc, a high performance moulding compound.
\newblock {\em SAMPE European conference}, 2001.

\bibitem{Aubry2001}
J~Aubry.
\newblock Hexmc—bridging the gap between prepreg and smc.
\newblock {\em Reinforced Plastics}, 45(6):38--40, 2001.

\bibitem{Visweswaraiah2018}
SB~Visweswaraiah, M~Selezneva, L~Lessard, and P~Hubert.
\newblock Mechanical characterisation and modelling of randomly oriented strand
  architecture and their hybrids--a general review.
\newblock {\em Journal of Reinforced Plastics and Composites}, 37(8):548--580,
  2018.

\bibitem{Ko2019a}
S~Ko, J~Yang, ME~Tuttle, and M~Salviato.
\newblock Effect of the platelet size on the fracturing behavior and size
  effect of discontinuous fiber composite structures.
\newblock {\em Composite Structures}, 227:111245, 2019.

\bibitem{Ko2019b}
S~Ko, J~Davey, S~Douglass, J~Yang, ME~Tuttle, and M~Salviato.
\newblock Effect of the thickness on the fracturing behavior of discontinuous
  fiber composite structures.
\newblock {\em Composites Part A: Applied Science and Manufacturing},
  125:105520, 2019.

\bibitem{troy22}
Troy Nakagawa, Seunghyun Ko, Cory Slaughter, Talal Abdullah, Guy Houser, and
  Marco Salviato.
\newblock Effects of aging on the mechanical and fracture properties of chopped
  fiber composites made from repurposed aerospace prepreg scrap and waste.
\newblock {\em Sustainable Materials and Technologies}, page e00470, 2022.

\bibitem{Fera2009a}
P~Feraboli, E~Peitso, F~Deleo, T~Cleveland, and PB~Stickler.
\newblock Characterization of prepreg-based discontinuous carbon fiber/epoxy
  systems.
\newblock {\em Journal of reinforced plastics and composites},
  28(10):1191--1214, 2009.

\bibitem{Fera2009b}
P~Feraboli, E~Peitso, T~Cleveland, and PB~Stickler.
\newblock Modulus measurement for prepreg-based discontinuous carbon
  fiber/epoxy systems.
\newblock {\em Journal of composite materials}, 43(19):1947--1965, 2009.

\bibitem{Baz1998}
ZP~Ba{\v{z}}ant and J~Planas.
\newblock {\em Fracture and size effect in concrete and other quasibrittle
  materials}.
\newblock CNC Press, 1998.

\bibitem{Sal21a}
ZP~Ba{\v{z}}ant, J~Le, and M~Salviato.
\newblock {\em Quasibrittle Fracture Mechanics and Size Effect. A First
  Course}.
\newblock Oxford University Press, 2021.

\bibitem{Baz2004}
ZP~Ba{\v{z}}ant.
\newblock Probability distribution of energetic-statistical size effect in
  quasibrittle fracture.
\newblock {\em Probabilistic engineering mechanics}, 19(4):307--319, 2004.

\bibitem{Baz2017}
ZP~Ba{\v{z}}ant and J~Le.
\newblock {\em Probabilistic Mechanics of Quasibrittle Structures: Strength,
  Lifetime, and Size Effect}.
\newblock Cambridge University Press, 2017.

\bibitem{Yao2020a}
Y~Qiao and M~Salviato.
\newblock Micro-computed tomography analysis of damage in notched composite
  laminates under multi-axial fatigue.
\newblock {\em Composites Part B: Engineering}, 187:107789, 2020.

\bibitem{Yao2018a}
CH~Mefford, Y~Qiao, and M~Salviato.
\newblock Failure behavior and scaling of graphene nanocomposites.
\newblock {\em Composite Structures}, 176:961--972, 2017.

\bibitem{Yao2018b}
Y~Qiao and M~Salviato.
\newblock Study of the fracturing behavior of thermoset polymer nanocomposites
  via cohesive zone modeling.
\newblock {\em Composite Structures}, 220:127--147, 2019.

\bibitem{Yao2018c}
Y~Qiao and M~Salviato.
\newblock Strength and cohesive behavior of thermoset polymers at the
  microscale: A size-effect study.
\newblock {\em Engineering Fracture Mechanics}, 213:100--117, 2019.

\bibitem{Baz1996}
ZP~Ba{\v{z}}ant, IM~Daniel, and Z~Li.
\newblock Size effect and fracture characteristics of composite laminates.
\newblock {\em J Eng Mater Technol}, 118(3):317--324, 1996.

\bibitem{Sal2019}
M~Salviato, K~Kirane, ZP~Ba{\v{z}}ant, and G~Cusatis.
\newblock Mode {I} and {II} interlaminar fracture in laminated composites: a
  size effect study.
\newblock {\em Journal of Applied Mechanics}, 86(9), 2019.

\bibitem{Sal2016a}
M~Salviato, K~Kirane, SE~Ashari, ZP~Ba{\v{z}}ant, and G~Cusatis.
\newblock Experimental and numerical investigation of intra-laminar energy
  dissipation and size effect in two-dimensional textile composites.
\newblock {\em Composites Science and Technology}, 135:67--75, 2016.

\bibitem{Li21}
W~Li, Y~Qiao, J~Fenner, K~Warren, M~Salviato, ZP~Ba{\v{z}}ant, and Cusatis G.
\newblock Elastic and fracture behavior of three-dimensional ply-to-ply angle
  interlock woven composites: Through-thickness, size effect, and multiaxial
  tests.
\newblock {\em Composites Part C: Open Access}, 4:100098, 2021.

\bibitem{Sal2016b}
M~Salviato, VT~Chau, W~Li, ZP~Ba{\v{z}}ant, and G~Cusatis.
\newblock Direct testing of gradual postpeak softening of fracture specimens of
  fiber composites stabilized by enhanced grip stiffness and mass.
\newblock {\em Journal of Applied Mechanics}, 83(11), 2016.

\bibitem{Brockmann2023}
J~Brockmann and M~Salviato.
\newblock The gap test–effects of crack parallel compression on fracture in
  carbon fiber composites.
\newblock {\em Composites Part A: Applied Science and Manufacturing},
  164:107252, 2023.

\bibitem{Denos2014}
BR~Denos.
\newblock Ct scan analysis for the characterization of fiber orientation in
  long discontinuous fiber composite materials.
\newblock Master's thesis, Purdue University, 2014.

\bibitem{Denos2017a}
BR~Denos, SG~Kravchenko, and RB~Pipes.
\newblock Progressive failure analysis in platelet based composites using
  ct-measured local microstructure.
\newblock In {\em SAMPE Conference}, pages 1421--1435, 2017.

\bibitem{Kravchenko2017}
S~Kravchenko.
\newblock {\em Failure analysis in platelet molded composite systems. PhD
  Dissertation}.
\newblock PhD thesis, Purdue University, 2017.

\bibitem{Favaloro2018}
AJ~Favaloro, DE~Sommer, BR~Denos, and RB~Pipes.
\newblock Simulation of prepreg platelet compression molding: Method and
  orientation validation.
\newblock {\em Journal of Rheology}, 62(6):1443--1455, 2018.

\bibitem{Wan2016b}
Y~Wan, I~Straumit, J~Takahashi, and SV~Lomov.
\newblock Micro-ct analysis of internal geometry of chopped carbon fiber tapes
  reinforced thermoplastics.
\newblock {\em Composites Part A: Applied Science and Manufacturing},
  91:211--221, 2016.

\bibitem{Wan2018}
Y~Wan, I~Straumit, J~Takahashi, and SV~Lomov.
\newblock Micro-ct analysis of the orientation unevenness in randomly chopped
  strand composites in relation to the strand length.
\newblock {\em Composite Structures}, 206:865--875, 2018.

\bibitem{Wan2021}
Y~Wan, X~Zhang, Y~Sato, and J~Takahashi.
\newblock Stochasticity modeling and analysis of elastic modulus of randomly
  oriented strands.
\newblock {\em Advanced Composite Materials}, 30(3):286--306, 2021.

\bibitem{Martulli2019}
LM~Martulli, L~Muyshondt, M~Kerschbaum, S~Pimenta, SV~Lomov, and Y~Swolfs.
\newblock Carbon fibre sheet moulding compounds with high in-mould flow:
  Linking morphology to tensile and compressive properties.
\newblock {\em Composites Part A: Applied Science and Manufacturing},
  126:105600, 2019.

\bibitem{Fera2010}
P~Feraboli, T~Cleveland, P~Stickler, and JC~Halpin.
\newblock Stochastic laminate analogy for simulating the variability in modulus
  of discontinuous composite materials.
\newblock {\em Composites Part A: Applied Science and Manufacturing},
  41(4):557--570, 2010.

\bibitem{Tuttle2017}
K~Harban and ME~Tuttle.
\newblock Reducing certification costs of discontinuous fiber composite
  structures via stochastic modeling.
\newblock {\em U.S. Dept. of Transportation FAA}, 2017.

\bibitem{Sel2017}
M~Selezneva, S~Roy, S~Meldrum, L~Lessard, and A~Yousefpour.
\newblock Modelling of mechanical properties of randomly oriented strand
  thermoplastic composites.
\newblock {\em Journal of Composite Materials}, 51(6):831--845, 2017.

\bibitem{Ko2018a}
S~Ko, K~Chan, R~Hawkins, R~Jayaram, C~Lynch, R~El~Mamoune, M~Nguyen,
  N~Pekhotin, N~Stokes, D~Wu, J~Yang, ME~Tuttle, and M~Salviato.
\newblock Experimental and numerical characterization of the intra-laminar
  fracturing behavior in discontinuous fiber composite structures.
\newblock In {\em Proceedings of the 33th ASC Conference, Seattle, WA, USA},
  pages 24--26, 2018.

\bibitem{Ko2018b}
S~Ko, K~Chan, R~Hawkins, R~Jayaram, C~Lynch, R~El~Mamoune, M~Nguyen,
  N~Pekhotin, N~Stokes, D~Wu, J~Yang, ME~Tuttle, and M~Salviato.
\newblock Characterization and computational modeling of the fracturing
  behavior in discontinuous fiber composite structures.
\newblock In {\em SAMPE Conference. Long Beach, CA.}, 2018.

\bibitem{Ko2020}
S~Ko, J~Yang, ME~Tuttle, and M~Salviato.
\newblock Stochastic computational modeling of the fracturing behavior in
  discontinuous fiber composite structures.
\newblock In {\em SAMPE Conference. Virtual}, 2020.

\bibitem{Ko2021}
S~Ko, T~Nakagawa, Z~Chen, J~Davey, T~Abdullah, L~Kuklenski, E~Adams, M~Soja,
  C~Park, W~Avery, J~Yang, and M~Salviato.
\newblock Experimental and numerical investigations of stochastic thickness
  effects in discontinuous fiber composites.
\newblock In {\em American Society of Composites Conference. Virtual}, 2021.

\bibitem{kravchenko2019a}
SG~Kravchenko, DE~Sommer, BR~Denos, AJ~Favaloro, CM~Tow, WB~Avery, and B~Pipes.
\newblock Tensile properties of a stochastic prepreg platelet molded composite.
\newblock {\em Composites Part A: Applied Science and Manufacturing},
  124:105507, 2019.

\bibitem{Sommer2020a}
DE~Sommer, SG~Kravchenko, BR~Denos, AJ~Favaloro, and RB~Pipes.
\newblock Integrative analysis for prediction of process-induced,
  orientation-dependent tensile properties in a stochastic prepreg platelet
  molded composite.
\newblock {\em Composites Part A: Applied Science and Manufacturing},
  130:105759, 2020.

\bibitem{Shah2020}
SZH Shah, RS~Choudhry, and S~Mahadzir.
\newblock A new approach for strength and stiffness prediction of discontinuous
  fibre reinforced composites (dfc).
\newblock {\em Composites Part B: Engineering}, 183:107676, 2020.

\bibitem{Visweswaraiah2017}
SB~Visweswaraiah, L~Lessard, and P~Hubert.
\newblock Interlaminar shear behaviour of hybrid fibre architectures of
  randomly oriented strands combined with laminate groups.
\newblock {\em Composite Structures}, 176:823--832, 2017.

\bibitem{Li2019}
Y~Li and S~Pimenta.
\newblock Development and assessment of modelling strategies to predict failure
  in tow-based discontinuous composites.
\newblock {\em Composite Structures}, 209:1005--1021, 2019.

\bibitem{Henry2017}
J~Henry and S~Pimenta.
\newblock Semi-analytical simulation of aligned discontinuous composites.
\newblock {\em Composites Science and Technology}, 144:230--244, 2017.

\bibitem{Sekisui}
{Sekisui Aerospace. https://www.sekisuiaerospace.com/; 2021}.

\bibitem{Gom}
{GOM, Braunschweig, Germany. https://www.gom.com.; 2021}.

\bibitem{Cochran1977}
WG~Cochran.
\newblock {\em Sampling Techniques}.
\newblock New York: John Wiley $\&$ Sons, 1977.

\bibitem{CMH2002}
{Handbook, M. MIL-HDBK-17: Composite materials handbook. Virginia: US
  Department of Defense, 2002.}

\bibitem{NSI}
{North Star Imaging. https://4nsi.com/; 2021}.

\bibitem{VGStudio2015}
{Volume Graphics GMBH, VGStudio MAX 3.0, Heidelberg, 2015.}

\bibitem{Denos2017b}
BR~Denos.
\newblock {\em Fiber Orientation Measurement in Platelet-Based Composites via
  Computed Tomography Analysis}.
\newblock PhD thesis, Purdue University, 2017.

\bibitem{Shapiro2011}
LG~Shapiro and GC~Stockman.
\newblock {\em Computer Vision}.
\newblock Pearson, 2011.

\bibitem{Steiner1994}
C~Dai and Steiner PR.
\newblock Spatial structure of wood composites in relation to processing and
  performance characteristics. part 2. modelling and simulation of
  randomly-formed flake layer network.
\newblock {\em Wood Sci Technol}, 28:135--146, 1994.

\bibitem{Sommer2020b}
DE~Sommer, SG~Kravchenko, and RB~Pipes.
\newblock A numerical study of the meso-structure variability in the compaction
  process of prepreg platelet molded composites.
\newblock {\em Composites Part A: Applied Science and Manufacturing},
  138:106010, 2020.

\bibitem{Tucker1987}
SG~Advani and CL~Tucker.
\newblock The use of tensors to describe and predict fiber orientation in short
  fiber composites.
\newblock {\em Journal of Rheology}, 31(751), 1987.

\bibitem{Kanit2003}
T~Kanit, S~Forest, I~Galliet, V~Mounoury, and D~Jeulin.
\newblock Determination of the size of the representative volume element for
  random composites: statistical and numerical approach.
\newblock {\em International Journal of solids and structures},
  40(13-14):3647--3679, 2003.

\bibitem{Abaqus}
{Dassault Systemes ABAQUS. ABAQUS Documentation, Providence, RI. 2018.}

\bibitem{Green2007}
BG~Green, MR~Wisnom, and SR~Hallett.
\newblock An experimental investigation into the tensile strength scaling of
  notched composites.
\newblock {\em Composites Part A: Applied Science and Manufacturing},
  38(3):867--878, 2007.

\bibitem{Jeremy22}
Jeremy Brockmann and Marco Salviato.
\newblock The gap test: Effects of crack parallel compression on fracture in
  carbon fiber composites.
\newblock {\em arXiv preprint arXiv:2207.12649}, 2022.

\bibitem{Baz1984}
ZP~Ba{\v{z}}ant.
\newblock Size effect in blunt fracture: concrete, rock, metal.
\newblock {\em Journal of engineering mechanics}, 110(4):518--535, 1984.

\bibitem{Baz1990}
ZP~Ba{\v{z}}ant and MT~Kazemi.
\newblock Determination of fracture energy, process zone longth and brittleness
  number from size effect, with application to rock and conerete.
\newblock {\em International Journal of fracture}, 44(2):111--131, 1990.

\bibitem{Sal2011}
M~Salviato, M~Zappalorto, and M~Quaresimin.
\newblock The effect of surface stresses on the critical debonding stress
  around nanoparticles.
\newblock {\em International journal of fracture}, 172(1):97--103, 2011.

\bibitem{Sal2013}
M~Salviato, M~Zappalorto, and M~Quaresimin.
\newblock Nanoparticle debonding strength: a comprehensive study on interfacial
  effects.
\newblock {\em International Journal of Solids and Structures},
  50(20-21):3225--3232, 2013.

\bibitem{Hashin1980}
Z~Hashin.
\newblock Failure criteria for unidirectional fiber composites.
\newblock {\em J Appl Mech}, 47:329--34, 1980.

\bibitem{Lapczyk2007}
I~Lapczyk and JA~Hurtado.
\newblock Progressive damage modeling in fiber-reinforced materials.
\newblock {\em Composites Part A: Applied Science and Manufacturing},
  38(11):2333--2341, 2007.

\bibitem{Baz1983}
ZP~Ba{\v{z}}ant and BH~Oh.
\newblock Crack band theory for fracture of concrete.
\newblock {\em Mat{\'e}riaux et construction}, 16(3):155--177, 1983.

\bibitem{Beghini2008a}
G~Cusatis, A~Beghini, and ZP~Ba{\v{z}}ant.
\newblock Spectral stiffness microplane model for quasibrittle composite
  laminates—part i: theory.
\newblock {\em Journal of Applied Mechanics}, 75(2), 2008.

\bibitem{Beghini2008b}
A~Beghini, G~Cusatis, and ZP~Ba{\v{z}}ant.
\newblock Spectral stiffness microplane model for quasibrittle composite
  laminates—part ii: calibration and validation.
\newblock {\em Journal of Applied Mechanics}, 75(2), 2008.

\bibitem{Sal2016c}
M~Salviato, SE~Ashari, and G~Cusatis.
\newblock Spectral stiffness microplane model for damage and fracture of
  textile composites.
\newblock {\em Composite Structures}, 137:170--184, 2016.

\bibitem{xue2022cylindrical}
Jing Xue and Kedar Kirane.
\newblock Cylindrical microplane model for compressive kink band failures and
  combined friction/inelasticity in fiber composites i: Formulation.
\newblock {\em Composite Structures}, 289:115382, 2022.

\bibitem{sabounchi2022microplane}
Saeed Sabounchi and Ferhun~C Caner.
\newblock Microplane model of cylindrical geometry for transversely isotropic
  polymer composites.
\newblock {\em Computers \& Structures}, 268:106807, 2022.

\bibitem{Camanho2002}
PP~Camanho and CG~D{\'a}vila.
\newblock Mixed-mode decohesion finite elements for the simulation of
  delamination in composite materials.
\newblock 2002.

\bibitem{Benzeggagh1996}
ML~Benzeggagh and M~Kenane.
\newblock Measurement of mixed-mode delamination fracture toughness of
  unidirectional glass/epoxy composites with mixed-mode bending apparatus.
\newblock {\em Composites science and technology}, 56(4):439--449, 1996.

\bibitem{kravchenko2019b}
SG~Kravchenko, DE~Sommer, BR~Denos, WB~Avery, and B~Pipes.
\newblock Structure-property relationship for a prepreg platelet molded
  composite with engineered meso-morphology.
\newblock {\em Composite Structures}, 210:430--445, 2019.

\bibitem{Turon2007}
A~Turon, CG~D\'{a}vila, PP~Camanho, and J~Costa.
\newblock An engineering solution for mesh size effects in the simulation of
  delamination using cohesive zone models.
\newblock {\em Engineering fracture mechanics}, 74(10):1665--1682, 2007.

\bibitem{McKay1980}
MD~McKay, RJ~Beckman, and WJ~Conover.
\newblock A comparison of three methods for selecting values of input variables
  in the analysis of output from a computer code.
\newblock {\em Technometrics}, 42(1):55--61, 2000.

\bibitem{Walpole1993}
RE~Walpole, RH~Myers, SL~Myers, and K~Ye.
\newblock {\em Probability and statistics for engineers and scientists},
  volume~5.
\newblock Macmillan New York, 1993.

\bibitem{Weibull1951}
W~Weibull.
\newblock A statistical distribution function of wide applicability.
\newblock {\em Journal of applied mechanics}, 18(3):293--297, 1951.

\bibitem{Amidror2002}
I~Amidror.
\newblock Scattered data interpolation methods for electronic imaging systems:
  a survey.
\newblock {\em Journal of electronic imaging}, 11(2):157--176, 2002.

\bibitem{Mindess2003}
S~Mindess, JF~Young, and D~Darwin.
\newblock {\em Concrete}.
\newblock Prentice Hall, 2003.

\end{thebibliography}
\clearpage

%\listoffigures  %\tableofcontents
%\listoftables

%%%%%%%%%%%%%%%%%%%%%%%%%%%%%%%%%%%%%%%%%%%%%%%%%%%%%%
%% Figures and Tables 
%%%%%%%%%%%%%%%%%%%%%%%%%%%%%%%%%%%%%%%%%%%%%%%%%%%%%%
%%% Use Times New Roman as a font inside of the figure.

% Figure Stress-Strain curves

% % Figure 2
% \begin{figure}
%  \centering
%   \includegraphics[trim=0cm 4cm 0cm 1cm, clip=true,width = 1\textwidth, scale=1]{figures/f-TestSigEpsCurves.pdf} \caption{\label{f-TestSigEpsCurves} } 
% \end{figure}

% Figure Fracture Surface

% Figure: Narrow, Square Modulus vs. Thickness

% Figure 5

% Figure 6

% Figure 7

% Figure 8

% Figure 9

% Figure 10

% Figure 11

% Figure 12

% Figure 13

% Figure 14

% Figure 15

% Figure 16

% Figure 17

% Figure 18

%% Figure sample for Appendix %%

% \renewcommand{\thefigure}{A\arabic{figure}}
% \setcounter{figure}{0}

% \begin{figure}
%  \centering
%   \includegraphics[trim=0cm 0cm 0cm 0cm, clip=true,width = 1\textwidth, scale=1]{} \caption{\label{f-FEMGeo} \sf Figure sample for appendix} 
% \end{figure}

% \renewcommand{\thefigure}{B\arabic{figure}}
% \setcounter{figure}{0}

\clearpage

% %%%%%%%%%%%%%%%%%%%%%%%%%%%%%%%%%%%%%%%%%%%%
% %%%%%%%% %%%% %%%%  Table %%%%%%%% %%%% %%%% 
% %%%%%%%%%%%%%%%%%%%%%%%%%%%%%%%%%%%%%%%%%%%%
\begingroup
\setlength\extrarowheight{15pt}

\begin{table}[ht]
\centering  % used for centering table
\caption{A summary of the experiment results normalized against the theoretical quasi-isotropic laminate modulus and strength made with the same prepreg system}
\begin{adjustbox}{max width=\textwidth}
\begin{tabular}{ c c c c c c c c} % centered columns
\toprule
%   \rule{0pt}{4ex}
 Platelet & Thickness [mm] & Normalized & Modulus & Normalized & Strength & Tested\\
 Size & (layers) & Modulus & CoV [\%] & Strength & CoV [\%] & Coupons\\[1mm]
 \hline
Narrow & $1.65~(11)$ & $0.907$ & $11.6$ & $0.366$ & $18.7$ & $17$\\
 & $3.81~(27)$ & $0.897$ & $12.9$ & $0.414$ & $13.6$ & $33$\\
 & $6.35~(45)$ & $0.938$ & $9.3$ & $0.443$ & $11.6$ & $33$\\[2mm]
 
 \hline
 
Square & $1.65~(11)$ & $0.781$ & $11.1$ & $0.223$ & $10.8$ & $15$\\
 & $3.81~(27)$ & $0.830$ & $6.7$ & $0.330$ & $8.3$ & $30$\\[1mm]
\bottomrule

\end{tabular}
\end{adjustbox}
% \begin{tablenotes}
%       \small
%       \item *Normalized against the theoretical modulus and strength of a quasi-isotropic laminate made with identical prepreg material properties.
%     \end{tablenotes}

\label{T1}
\end{table}
\endgroup
%%%%%%%%%%%%%%%%%%%%%%%%%%%%%%%%%%%%%%%%%%%%%%%%%%%%%%%%

% Table 2 %%%%%%%%%%%%%%%%%%%%%%%%%%%%%%%%%%%%%%%%%%%%%%%%%%%%%%%
\setlength\extrarowheight{15pt}
\begin{table}[ht]
\centering  % used for centering table
\caption{A summary of the normalized experiment B-basis design values}
\begin{adjustbox}{max width=\textwidth}
\begin{tabular}{ c c c c} % centered columns

\toprule
%   \rule{0pt}{4ex}
 Platelet   & Thickness [mm] & Normalized modulus    & Normalized strength\\
 Size       & (layers)      & B-basis value         & B-basis value \\[1mm]
 \hline
Narrow & $1.65~(11)$ & $0.685$ & $0.214$\\
 & $3.81~(27)$ & $0.684$ & $0.315$ \\
 & $6.35~(45)$ & $0.748$ & $0.338$ \\[2mm]
 \hline%
Square & $1.65~(11)$ & $0.595$ & $0.163$ \\
 & $3.81~(27)$ & $0.723$ & $0.268$ \\[1mm]
\bottomrule

\end{tabular}
\end{adjustbox}
\label{T2}
\end{table}

%%%%%%%%%%%%%%%%%%%%%%%%%%%%%%%%%%%%%%%%%%%%%%%%%%%%%%%%
%%%%%%%%%%%%%%%% Table for the Appendix %%%%%%%%%%%%%%%%
%%%%%%%%%%%%%%%%%%%%%%%%%%%%%%%%%%%%%%%%%%%%%%%%%%%%%%%%
% \renewcommand{\thetable}{A\arabic{table}}
% \setcounter{table}{0}
% \begin{table}[ht]
% \centering  % used for centering table
% \caption{\sf TABLE SAMPEL 2}

% \begin{adjustbox}{max width=\textwidth}
% \begin{tabular}{ c c c} % centered columns (4 columns)
%  \hline
%   \rule{0pt}{4ex}
%  Properties &  T$700$G  & Matrix  \\ [1 ex]
%  \hline
% Platelet initial thickness, $t_p$ [mm] & $0.139$ & Varies\\
% Longitudinal modulus, $E_{1}$ [GPa] & $135$ & $3$ \\
% Transverse modulus, $E_{2}$ [GPa] & $10$ & $3$ \\
% Shear modulus, $G_{12}$ [GPa] & $5$ & $1.1$\\
% Poisson ratio, $\nu_{12}$ & $0.3$ & $0.35$ \\

% \hline%inserts single line
% \end{tabular}
% \end{adjustbox}
% \label{T3}
% \end{table}

% \clearpage

%%%%%%%%%%%%%%%%%%%%%%%%%%%%%%%%%%%%%%%%%%%%%%%%%%%%%%%%

\end{document}